\documentclass[journal]{IEEEtran}
\usepackage[dvips]{graphicx}
\usepackage{epsfig}
\usepackage{amsmath}
\usepackage{epsfig}
\usepackage{array}
\usepackage{amssymb}
\usepackage{color}
\newtheorem{Prop}{\textbf{Property}}

\begin{document}

\title{Fundamental Properties of Full-Duplex Radio for Secure Wireless Communications}

\author{Yingbo Hua\thanks{The authors are with
Department of Electrical and Computer Engineering,
University of California, Riverside, CA 92521, USA. Emails: yhua@ece.ucr.edu; qzhu005@ucr.edu; and rsohr001@ucr.edu. This work was partially presented in a Keynote at the Full-Duplex Technology Workshop of IEEE PIMRC-2017, Montreal, Canada, Oct 10, 2017. The research was supported in part by the Army Research Office under Grant Number W911NF-17-1-0581. The views and conclusions contained in this
document are those of the authors and should not be interpreted as representing the official policies, either
expressed or implied, of the Army Research Office or the U.S. Government. The U.S. Government is
authorized to reproduce and distribute reprints for Government purposes notwithstanding any copyright
notation herein.},
Qiping Zhu, Reza Sohrabi
}

\maketitle

\begin{abstract}
This paper presents a number of fundamental properties of full-duplex radio for secure wireless communication under some simple and practical conditions. In particular, we consider the fields of secrecy capacity of a wireless channel between two single-antenna radios (Alice and Bob) against an unknown number of single-antenna eavesdroppers (Eves) from unknown locations, where Alice and Bob have zero knowledge (except a model) of the large-scale-fading channel-state-information of Eves. These properties show how the secrecy capacity is distributed in terms of the location of any Eve, how the optimal jamming power applied by the full-duplex radio varies with various parameters, and how bad or good the worst cases are. In particular, these properties show how the quality of self-interference cancelation/suppression affects various aspects of the fields of secrecy capacity. The cases of colluding Eves and non-colluding Eves are treated separately and yet coherently. For non-colluding Eves, asymptotically constant fields of secrecy capacity are revealed. For each of the two cases, we also treat subcases with or without small-scale fading.
\end{abstract}
\section{Introduction}
Wireless communication\footnote{``Wireless communication'' and ``wireless communications'' are considered interchangeable.} is already firmly embedded in people's lives around the world. Secure wireless communication is important for all of us from individuals and families to institutions and governments. Yet, the physical medium for wireless communication is intrinsically open to all wireless devices in any given space, time and bandwidth. This makes wireless communication particularly vulnerable to eavesdropping.

For secure communication, cryptography at the network and upper layers is an efficient,  effective and indispensable tool \cite{Ferguson2010}. Cryptography also makes wireless communication highly secure against eavesdropping as long as secret keys are kept secure and renewed frequently. However, when a secret key itself is distributed through wireless channels, the physical layer security becomes essential for secure wireless communications.
 A recent survey on physical layer security is available in \cite{Mukherjee2014}, and a more general survey on wireless security is shown in \cite{Zou2016}.

In this paper, we are interested in physical layer security via the use of full-duplex radio. A full-duplex radio is able to transmit and receive at the same time and same frequency, which differs from the conventional radio which only transmits and receives, respectively, in two separate time slots or at two separate frequencies. While research continues in order to improve the quality of full-duplex radio, prototypes of full-duplex radio can be found in \cite{Krish2016} and the references therein.

A full-duplex radio is uniquely equipped for secure wireless communication. As a full-duplex radio receives a secret key from another radio, it can also transmit jamming noise\footnote{The idea of using radio jamming to prevent adversaries from receiving secret information dates back at least to the era of World War II.} at the same time and same frequency to prevent eavesdroppers (Eves) from receiving the same key. It is this intrinsic characteristic of full-duplex radio that has recently drawn much interest in exploiting full-duplex radio for secure wireless communication. Such examples include fast power allocation for both a transmitter and a jamming receiver in a multicarrier setting \cite{Chen2017}, utilization of full-duplex for secure decentralized wireless network \cite{TXZheng2017}, resource allocation  \cite{Abedi2017} and a hierarchical game \cite{Tang2017} against an active full-duplex Eve \cite{Tang2017}, optimal power allocation for a multiuser MISO network against multiple Eves \cite{Akgun2017}, and an early effort where MIMO full-duplex is exploited \cite{Zheng2013}. Many more relevant works can be found from the references therein.

However, a common assumption made in all these prior works is that the legitimate radios (such as Alice and Bob) know exactly how many Eves are nearby and also the large-scale-fading channel-state-information (L-CSI) of all Eves. This assumption is difficult to justify. If an Eve is passive and hidden (as it is often the case), it is virtually impossible for another radio to know its presence and let alone its L-CSI. Even if an Eve is active, its L-CSI is still hard to estimate without knowing either its location or its transmitted power. In other words, for most realistic situations, there is no way for a legitimate radio to know how many Eves are nearby, where they are, or what their L-CSI is.

In this paper, we focus on two basic schemes for key distribution (or any secret information\footnote{If jamming is mainly used for exchange of secret keys, it would not overly interfere the rest of the network.}) between two single-antenna radios (called Alice and Bob) in the presence of single-antenna Eves, where we do not assume that Alice and Bob have any knowledge of the number of Eves, their locations or their L-CSI. We will adopt a reasonable model of L-CSI to map between  L-CSI of any possible Eve and its approximate location. For small-scale-fading channel-state-information (S-CSI), we will use the Rayleigh-amplitude statistical model as is commonly used and well justified for scattering-rich mobile environment. Based on the above models, we will study the field of the secrecy capacity of Alice and Bob, which is a function of the location of Eve. If the field is positive everywhere in the space where Eves could reside, then there is a positive secrecy capacity against any and all Eves in the space.

The primary findings of this work will be highlighted in a list of properties and, where effective, by illustrative figures. Some of these properties may be more important than others, some may be more surprising than others, and some of the proofs may be more straightforward than others. But we feel that they all provide fundamental insights into how the secrecy capacity of the studied system is distributed in terms of the (unknown) location of Eve, what and where the worst cases are, and how the jamming power from a full-duplex radio and the quality of its self-interference cancellation/suppression affect the fields of secrecy capacity.

This paper should be easy to read since the mathematical tools used in this paper are little more than what a college-level training in mathematics and statistics provides for engineering students. All proofs that are tedious manipulations are omitted. But for all results, we will at least provide the right directions (if not obvious) for readers to verify by themselves.

For the rest of this paper, we will start with mathematical normalization of the problem in section \ref{sec:normalization} where we remove all redundant variables that are easy to deal with whenever needed. We will treat the case of colluding Eves in section \ref{sec:collusion} and the case of non-colluding Eves in section \ref{sec:no_collusion}. The work in section \ref{sec:collusion} is also very much coherent with that in section \ref{sec:no_collusion} although the two cases are totally different scenarios in applications. In each of the two sections, we first handle the situation without small-scale fading and then the situation with small-scale fading. The final remarks are given in section \ref{sec:final}.

\section{Mathematical Normalization}\label{sec:normalization}
If Alice uses the power $P_T'$ to transmit a key to Bob, and Bob (a full-duplex radio) receives the key and also sends out a jamming noise of the power $P_J'$, then the channel capacity\footnote{All capacity expressions in this paper have the unit in bits per channel use or equivalently in bits per second per Hertz.} from Alice to Bob usable for the packet of the key is known (e.g., see \cite{Tse2005}) to be
\begin{equation}\label{eq:CAB}
  C_{A,B}=\log_2(1+SNR_{A,B} )
\end{equation}
 with
$
SNR_{A,B} =\frac{g'P_T'}{P_{N,B}'+\rho'P_J'}
$
where $g'$ is the squared amplitude of the (actual) channel gain from Alice to Bob, $P_{N,B}'$ is the (actual) variance of the background noise at Bob, and $\rho'$ is the squared amplitude of the (actual) residual self-interference (SI) channel gain of Bob. The residual SI channel gain results from a combined effect of SI suppression at all stages, including antenna SI isolation, radio-frequency front-end SI cancellation and baseband SI cancellation. For principles of SI cancellation, see \cite{Hua2015}, for example, and the references therein.

At the same time, the channel capacity from Alice to Eve is
\begin{equation}\label{eq:CAE}
  C_{A,E}=\log_2(1+SNR_{A,E})
\end{equation}
with
$
  SNR_{A,E} =\frac{a'P_T'}{P_{N,E}'+b'P_J'}
$
where $a'$ is the squared amplitude of the (actual) channel gain from Alice to Eve, $b'$ is the squared amplitude of the (actual) channel gain from Bob to Eve, and $P_{N,E}'$ is the (actual) variance of the background noise at Eve.

To remove all redundant variables, we will use the following normalized variables: $P_T\doteq\frac{g'P_T'}{P_{N,B}'}$, $P_J\doteq\frac{g'P_J'}{P_{N,B}'}$, $\rho\doteq\frac{\rho'}{g'}$, $a\doteq\frac{a'P_{N,B}'}{g'P_{N,E}'}$ and $b\doteq\frac{b'P_{N,B}'}{g'P_{N,E}'}$, which are uniquely corresponding to $P_T'$, $P_J'$, $\rho'$, $a'$ and $b'$, respectively, as long as Alice and Bob know  the actual channel gain between them and the actual noise variances at all nodes. 
 For convenience, we will assume that the variance of the background noise\footnote{which includes  thermal noise from within the device and radio noise from numerous sources of both man-made and nature-made} is the same for all nodes.

 With the normalized variables, we can rewrite SNRs in \eqref{eq:CAB} and \eqref{eq:CAE} as follows\footnote{With zero impact on reading, we will choose to ignore adding the period ``$.$'' at the end of an equation line that is also the end of a sentence.}:
\begin{equation}\label{eq:SNRAB}
  SNR_{A,B} =\frac{P_T}{1+\rho P_J}
\end{equation}
\begin{equation}\label{eq:SNRAE}
  SNR_{A,E} =\frac{aP_T}{1+bP_J}
\end{equation}
 We will use the expressions of \eqref{eq:SNRAB} and \eqref{eq:SNRAE} for the case without small-scale fading. In the case with small-scale fading, we will make modifications later.

\begin{figure}
  \centering
  \includegraphics[width=4.5cm]{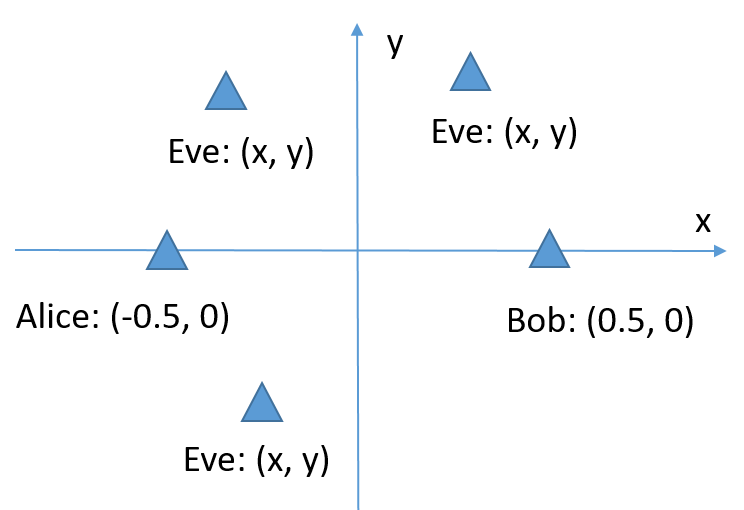}\\
  \caption{Normalized coordinates.}\label{Coordinates}
\end{figure}

 Because of the above normalization, we will use the following  terminology without loss of generality:
\begin{enumerate}
 \item The distance between Alice and Bob is one, and the large-scale fading gain between them is one. Both nodes are located on the $x$-axis of a two-dimensional plane\footnote{There is no loss of generality here since the 2-D results in this paper can be easily mapped into the 3-D space by a rotation around the $x$-axis.}: Alice is at $(-0.5,0)$, and Bob is at $(0.5,0)$. Eve's location is denoted by $(x,y)$. See Fig. \ref{Coordinates}.
  \item The large-scale fading factor from Alice to Eve is denoted by $a=\frac{1}{d_A^\alpha}$ with $d_A=\sqrt{(x+0.5)^2+y^2}$, and the large-scale fading factor from Bob to Eve is denoted by $b=\frac{1}{d_B^\alpha}$ with $d_B=\sqrt{(x-0.5)^2+y^2}$. Here, $\alpha\geq 2$ is the path loss exponent. 
  \item The residual SI channel gain\footnote{For convenience, we will also refer to ``squared amplitude of channel gain'' as ``channel gain'' unless clarification is needed.} of a full-duplex radio is $\rho$. If $\rho$ ($=\frac{\rho'}{g'}$) is less than one, it means that the actual amount of self-interference suppression used in the full-duplex radio is more than the large-scale path loss from Alice to Bob. It is important to remember that for a full-duplex radio with a fixed and actual residual SI channel gain $\rho'$,  $\rho$ can be larger or smaller than one depending on the actual channel gain $g'$ between Alice and Bob.
  \item The transmitted power\footnote{We assume that $P_T$ is strictly larger than zero because if $P_T=0$ there would be no transmission of information.} from Alice is $P_T>0$. The jamming power from Bob is $P_J\geq 0$. And the noise variance at all nodes is one.
\end{enumerate}

The rest of this paper is divided into two cases: colluding Eves and non-colluding Eves. For each case, we also consider two subcases with or without small-scale fading.

\section{With collusion among eavesdroppers}\label{sec:collusion}
We do not need to assume that Alice or Bob knows the number of Eves or the locations of Eves. But in this section, we assume that there could be collusion among all Eves at unknown locations. This means that if one Eve at one location steals a key, then all other Eves at any locations may also know the key.

\subsection{Without small-scale fading}
The secrecy capacity of the channel from Alice to Bob against Eve at location $(x,y)$ is known (e.g., see \cite{Mukherjee2014}) to be
\begin{equation}\label{eq:SABxy}
  S_{A,B,x,y}= [C_{A,B}-C_{A,E}]^+
\end{equation}
where $(x)^+=\max\{0,x\}$. Then, the secrecy capacity of the channel against all possible Eves is obviously the worst case:
\begin{equation}\label{}
  S_{A,B}=\min_{x,y}S_{A,B,x,y}
\end{equation}
We will see that without any constraint on $(x,y)$, $S_{A,B}$ would be always zero. However, it is practical to assume\footnote{Strictly speaking, this is a tiny amount of CSI about Eves. But this is negligible compared to the conventional definition of CSI.} that there is a radius $\Delta$ around Alice within which there is no Eve. Naturally, this radius could be small or large, depending on applications. Hence, we will define the secrecy capacity in this case as
\begin{equation}\label{}
  S_{A,B}=\min_{x,y, d_A\geq \Delta}S_{A,B,x,y}
\end{equation}
In order to understand this secrecy capacity, it is sufficient to understand how $S_{A,B,x,y}$ is distributed in terms of $(x,y)$, which is studied next.

Define a SNR ratio:
\begin{equation}\label{eq:SNRratio}
  \lambda_{x,y} \doteq \frac{SNR_{A,B}}{SNR_{A,E}}=\frac{1+bP_J}{a(1+\rho P_J)}
\end{equation}
Obviously, $S_{x,y}>0$ iff (if and only if) $\lambda_{x,y}>1$. In terms of $\lambda_{x,y}$, $S_{A,B,x,y}$ has the following expression:
\begin{equation}\label{}
  S_{A,B,x,y}=\left \{\begin{array}{cc}
  (\log_2e)\left (1-\frac{1}{\lambda_{x,y}}\right )^+ SNR_{A,B}, & P_T\rightarrow 0\\
                    \left ( \log_2 \lambda_{x,y}\right )^+, & P_T\rightarrow \infty
                  \end{array}
   \right .
\end{equation}
The following property shows a deeper insight into the conditions under which the secrecy\footnote{For convenience, secrecy capacity is also referred to as secrecy.} $S_{A,B,x,y}$ is positive.
\begin{Prop}\label{Prop1}
Let $\gamma=\frac{a-1}{b-\rho a}$. Then,
$S_{A,B,x,y}>0$ iff
  \begin{enumerate}
    \item $b-\rho a>0$ and $P_J>\gamma$; or
    \item $b-\rho a<0$ and $P_J<\gamma$; or
    \item $b-\rho a=0$ and $a<1$.
  \end{enumerate}
\end{Prop}
\begin{IEEEproof}
The proof is straightforward based on the statement following \eqref{eq:SNRratio}.
\end{IEEEproof}

The condition $b-\rho a>0$ means that $\rho$ must be small enough for the given $a$ and $b$ while $b-\rho a<0$ requires $\rho$ to be large enough. Keep in mind that both $a$ and $b$ are strictly positive and depend on Eve's location $(x,y)$. Also, iff $a>1$, $d_A<1$ (Eve is inside the unit circle around Alice); and iff $b<1$,  $d_B>1$ (Eve is outside the unit circle around Bob). In order to view Property \ref{Prop1} geometrically, let us define the following four regions:
\begin{enumerate}
 \item Region $\mathcal{R}_1$: $b-\rho a>0$ and $a<1$.
  \item Region $\mathcal{R}_2$: $b-\rho a>0$ and $a\geq 1$.
  \item Region $\mathcal{R}_3$: $b-\rho a\leq 0$ and $a<1$. This region vanishes\footnote{i.e., the two inequalities do not hold at the same time.} iff $\rho\leq \frac{1}{2^\alpha}$.
  \item Region $\mathcal{R}_4$: $b-\rho a\leq 0$ and  $a\geq 1$.
\end{enumerate}
Then, Property \ref{Prop1} implies:
\begin{Prop}\label{Prop1a}
\begin{enumerate}
  \item If  $(x,y)\in\mathcal{R}_1$, then $S_{A,B,x,y}>0$ for any $P_J\geq 0$.
  \item If $(x,y)\in\mathcal{R}_2$, then $S_{A,B,x,y}>0$ iff $P_J>\gamma$.
  \item If $(x,y)\in\mathcal{R}_3$, then $S_{A,B,x,y}>0$ if $P_J=0$.
  \item If $(x,y)\in\mathcal{R}_4$, then $S_{A,B,x,y}=0$ for any $P_J\geq 0$.
\end{enumerate}
\end{Prop}
\begin{IEEEproof}
It follows from Property \ref{Prop1}.
\end{IEEEproof}

For $\mathcal{R}_3$, it will be shown later that $\arg\max_{P_J} S_{A,B,x,y}=0$, i.e., $P_J=0$ is optimal.
We see that unless Eve is in $\mathcal{R}_4$, there is $P_J\geq 0$ such that $S_{A,B,x,y}$ is positive. Therefore, in order to have an overall positive secrecy, i.e., $S_{A,B}>0$, it is necessary that $\mathcal{R}_4\subset \mathcal{R}_{d_A<\Delta}$ (i.e., $\mathcal{R}_4$ belongs to the region where $d_A<\Delta$). One can verify that for $\Delta\leq 1$, $\mathcal{R}_4\subset \mathcal{R}_{d_A<\Delta}$ iff $\rho<\frac{\Delta^\alpha}{(1+\Delta)^\alpha}$.  (Note that if $\Delta> 1$, then by definition of $\mathcal{R}_4$, $\mathcal{R}_4\subset \mathcal{R}_{d_A\leq 1}$ as always, and hence $\mathcal{R}_4\subset \mathcal{R}_{d_A<\Delta}$ for any $\rho$.) For example, if $\Delta=0.1$ and $\alpha=2$, then we need $\rho<0.008\approx-21$dB (i.e., SI suppression needs to be 21dB more than the path loss from Alice to Bob).

In order to visualize Property \ref{Prop1a}, we need to visualize the region $\mathcal{R}_{\rho}$ defined by $b-\rho a>0$:
\begin{Prop}\label{Prop_of_shape}
\begin{enumerate}
\item If $\rho=1$, then $b-\rho a>0$ is equivalent to $x>0$.
  \item If $\rho<1$,  then $b-\rho a>0$ is equivalent to
  \begin{equation}
  (x+x_0)^2+y^2>r_\rho^2
\end{equation}
with $x_0=\frac{1+\rho^{\frac{2}{\alpha}}}{2(1-\rho^{\frac{2}{\alpha}})}>\frac{1}{2}$ and
$r_\rho = \sqrt{x_0^2-\frac{1}{4}}$. That is, $\mathcal{R}_{\rho}$ is everywhere outside a circular disk in the left half plane, and Alice is inside the disk.
\item If $\rho>1$,  then $b-\rho a>0$ is equivalent to
  \begin{equation}
  (x+x_0)^2+y^2<r_\rho^2
\end{equation}
which means that $\mathcal{R}_{\rho}$ is now everywhere inside a circular disk in the right half plane, and Bob is inside this disk.
\end{enumerate}
\end{Prop}
\begin{IEEEproof}
The proof is straightforward.
\end{IEEEproof}

\begin{figure}
  \centering
  \includegraphics[width=4.5cm]{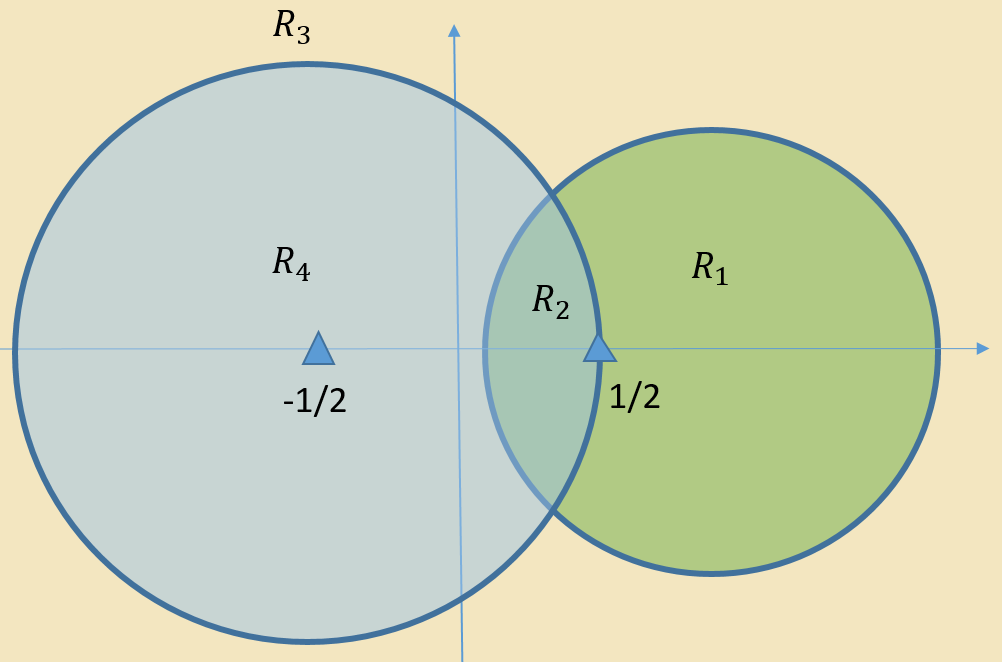}\\
  \caption{Illustration of $\mathcal{R}_1$, $\mathcal{R}_2$, $\mathcal{R}_3$ and $\mathcal{R}_4$ for $\rho>1$. }\label{regions_for_rho_larger_than_1}
\end{figure}

\begin{figure}
  \centering
  \includegraphics[width=4.5cm]{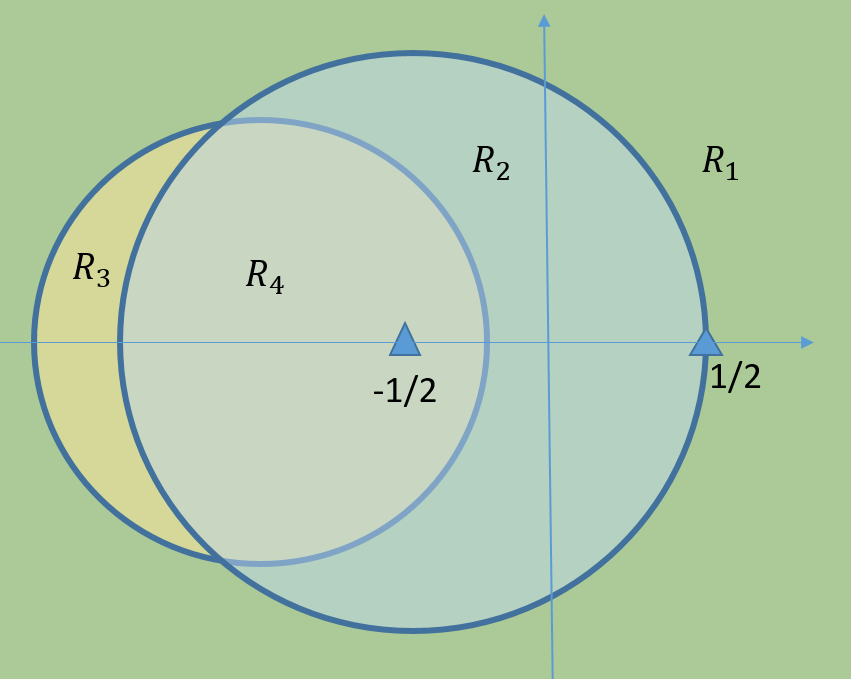}\\
  \caption{Illustration of $\mathcal{R}_1$, $\mathcal{R}_2$, $\mathcal{R}_3$ and $\mathcal{R}_4$ for $\rho<1$. $\mathcal{R}_3$ vanishes iff $\rho< \frac{1}{2^\alpha}$.}\label{regions_for_rho_less_than_1}
\end{figure}

\begin{figure}
  \centering
  \includegraphics[width=4.5cm]{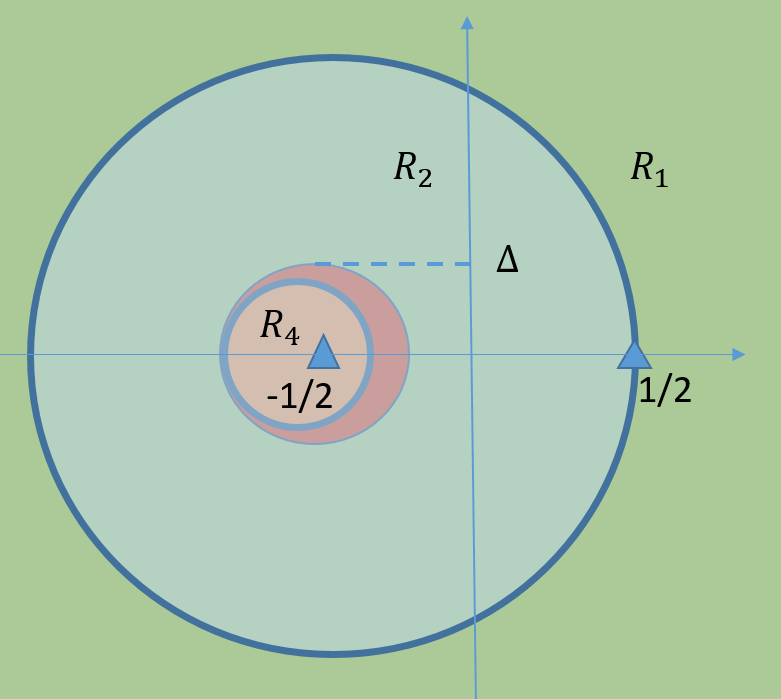}\\
  \caption{Illustration of $\mathcal{R}_4\subset \mathcal{R}_{d_A< \Delta}$ iff $\rho<\frac{\Delta^\alpha}{(1+\Delta)^\alpha}$ assuming $\Delta\leq 1$.}\label{region_4_inside_Delta}
\end{figure}

The four regions $\mathcal{R}_1$, $\mathcal{R}_2$, $\mathcal{R}_3$ and $\mathcal{R}_4$ are illustrated in Figs. \ref{regions_for_rho_larger_than_1}-\ref{regions_for_rho_less_than_1}. Illustrated in Fig. \ref{region_4_inside_Delta} is the case where $\mathcal{R}_4$ shrinks into $\mathcal{R}_{d_A<\Delta}$.

\begin{Prop}\label{Prop_of_ring}
Given $\rho<\frac{1}{2^\alpha}$, $P_J>0$ and $\mathcal{R}_{P_J}\doteq\{(x,y)\in \mathcal{R}_\rho|P_J\leq \gamma\}$, then  $S_{A,B,x,y}=0$ iff $(x,y)\in \mathcal{R}_4\cup\mathcal{R}_{P_J}$.
\end{Prop}
\begin{IEEEproof}
Under $\rho<\frac{1}{2^\alpha}$, $\mathcal{R}_4\subset \mathcal{R}_{d_A<1}$. If $(x,y)\in \mathcal{R}_4$, we know that $S_{A,B,x,y}=0$. If $(x,y)\notin \mathcal{R}_4$ but $(x,y)\in \mathcal{R}_{P_J}$, then we must have $b-\rho a>0$ and $a>1$ so that $P_J\leq \gamma$, which implies by Property \ref{Prop1} that $S_{A,B,x,y}=0$. If $(x,y)\notin \mathcal{R}_4\cup\mathcal{R}_{P_J}$, then   $S_{A,B,x,y}>0$ by Property \ref{Prop1a}.
\end{IEEEproof}

This property says that for a fixed $P_J>0$, there is an additional ring outside $\mathcal{R}_4$ where  $S_{A,B,x,y}=0$. This ring diminishes as $P_J$ becomes infinite.

\begin{Prop}\label{Prop_of_secrecy_distribution}
\begin{enumerate}
 \item Provided $S_{A,B,x,y}>0$, $S_{A,B,x,y}$ decreases if $a$ increases (Eve moves towards Alice) or $b$ decreases (Eve moves away from Bob).
  \item Let $\rho<\frac{\Delta^\alpha}{(1+\Delta)^\alpha}$ with $\Delta\leq 1$. Subject to $P_J>\gamma$,
  \[(x^*, y^*)\doteq\arg\min_{(x,y)\in\mathcal{R}_{d_A\geq\Delta}}S_{A,B,x,y}=(-\Delta-0.5,0)\] i.e., the most harmful location of Eve subject to $P_J>\gamma$ and $d_A\geq \Delta$ is at $\Delta$ distance to the left of Alice. Also note $(x^*, y^*) =  \arg \max_{(x,y)\in\mathcal{R}_{d_A\geq\Delta}} C_{A,E}
  =\arg \min_{(x,y)\in\mathcal{R}_{d_A\geq\Delta}} \lambda_{x,y} = \arg \min_{(x,y)\in\mathcal{R}_{d_A\geq\Delta}} \frac{b}{a}$.
\end{enumerate}
\end{Prop}
\begin{IEEEproof}
Part 1 is obvious from the definition of $S_{A,B,x,y}$ where $SNR_{A,B}$ is independent of $a$ and $b$, and $SNR_{A,E}$ is an increasing function of $a$ but a decreasing function of $b$. To prove Part 2, we first use Part 1 which suggests that for a fixed $a=\frac{1}{d_A^\alpha}$, the minimum of $S_{A,B,x,y}$ is achieved by the smallest $b$ which is $b=\frac{1}{(1+d_A)^\alpha}$. Then, subject to $d_A\geq \Delta$, $S_{A,B,x,y}$ is minimized by the above $a$ and $b$ with $d_A=\Delta$, which corresponds to $(x,y)=(-\Delta-0.5,0)$.
\end{IEEEproof}

\begin{figure}
  \centering
  \includegraphics[width=5cm]{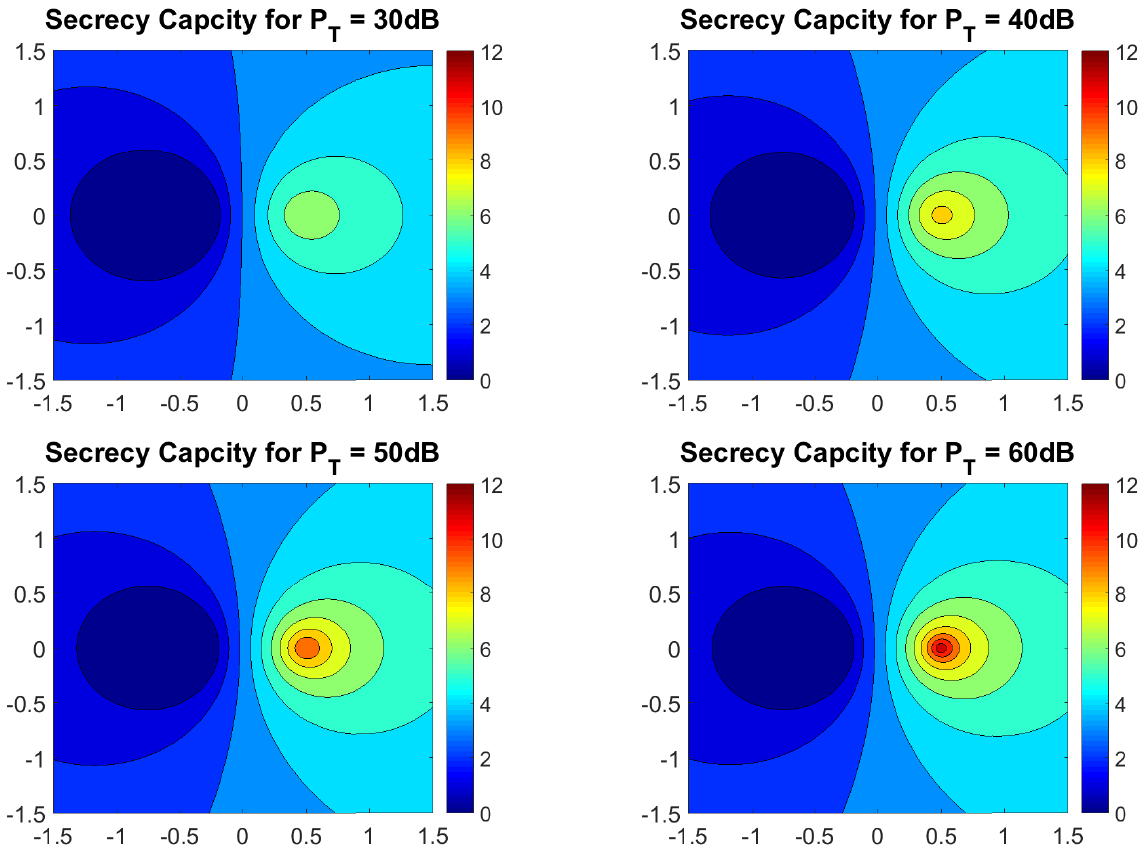}\\
  \caption{$S_{A,B,x,y}$ vs $(x,y)$ with $P_J=\sqrt{\frac{P_T}{\rho}}$, $\rho=0.1$ and $\alpha=2$.
  The darkest blue region is $\mathcal{R}_4\cup \mathcal{R}_{P_J}$.}\label{secrecy_capacity_single_constant_01}
\end{figure}

\begin{figure}
  \centering
  \includegraphics[width=5cm]{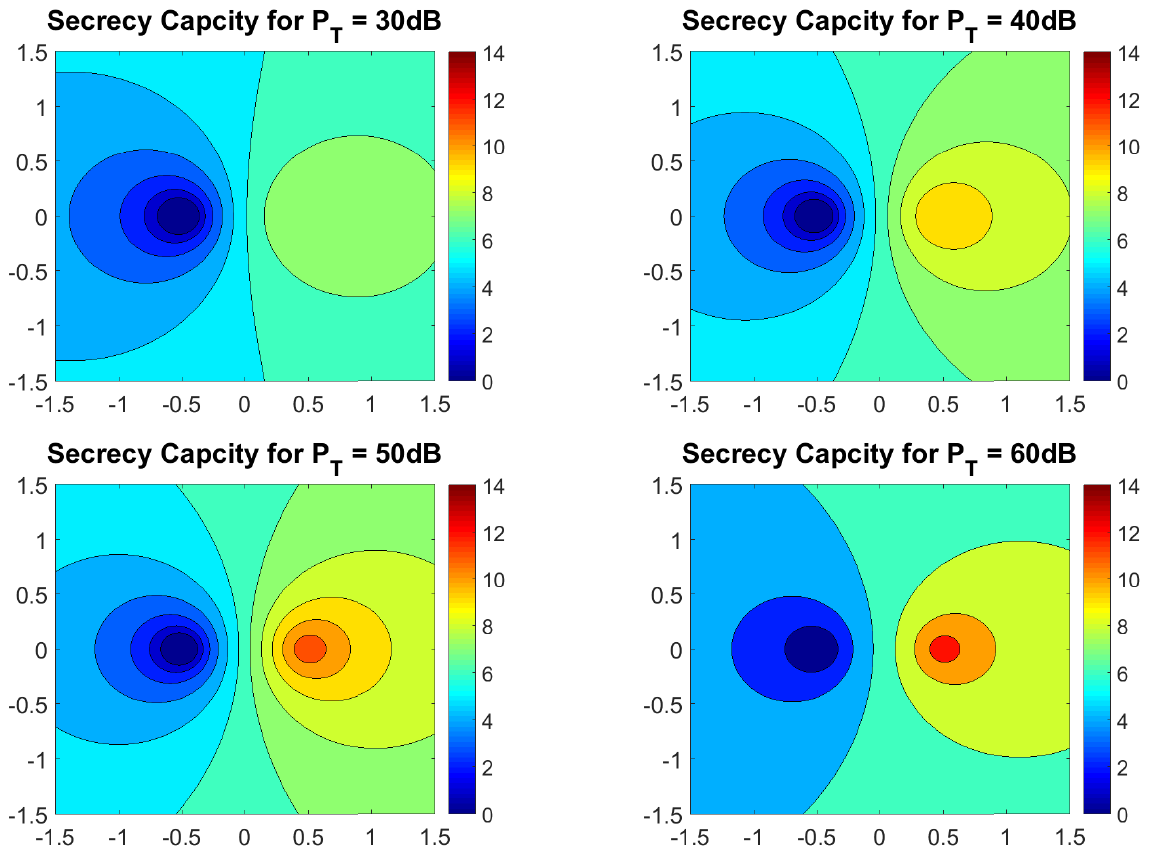}\\
  \caption{$S_{A,B,x,y}$ vs $(x,y)$ with $P_J=\sqrt{\frac{P_T}{\rho}}$, $\rho=0.01$ and $\alpha=2$.
  The darkest blue region is $\mathcal{R}_4\cup \mathcal{R}_{P_J}$.}\label{secrecy_capacity_single_constant_001}
\end{figure}

Shown in Figs. \ref{secrecy_capacity_single_constant_01}-\ref{secrecy_capacity_single_constant_001} is how $S_{A,B,x,y}$ is distributed over $(x,y)$ subject to $P_J=\sqrt{\frac{P_T}{\rho}}$. We will see that the choice of $P_J=\sqrt{\frac{P_T}{\rho}}$ is an asymptotical form of the optimal jamming power. In these two figures, the secrecy $S_{A,B,x,y}$ is zero in $\mathcal{R}_4\cup\mathcal{R}_{P_J}$ which is the darkest blue region.  We see that $\mathcal{R}_4\cup\mathcal{R}_{P_J}$ shrinks as $\rho$ becomes smaller.

\begin{Prop}\label{Prop_of_optimal_PJ}
Let $P_{J,opt,x,y} \doteq\arg\max_{P_J} S_{A,B,x,y}$. Then:
\begin{enumerate}
 \item For $(x,y)\in \mathcal{R}_1$,
     \begin{equation}\label{eq:optimal_PJ}
       P_{J,opt,x,y} =
  \left [\gamma+\sqrt{\gamma^2+\beta}\right ]^+\geq 0
     \end{equation}
with
$\gamma=\frac{a-1}{b-\rho a}$ and $\beta=\frac{ab-\rho +a P_T (b-\rho)}{\rho b(b-\rho a)}$.
\item For $(x,y)\in \mathcal{R}_2$, $P_{J,opt,x,y}$ is strictly positive and given by the equality in \eqref{eq:optimal_PJ}.
    \item For $(x,y)\in \mathcal{R}_3$, $P_{J,opt,x,y}=0$.
\end{enumerate}
\end{Prop}

\begin{IEEEproof}
One can verify that if $S_{A,B,x,y}>0$, then
\begin{equation}\label{}
  \frac{\partial S_{A,B,x,y}}{\partial P_J}=\frac{(\log_2 e) P_T}{g(P_J)}(-c_2P_J^2 +c_1 P_J + c_0)
\end{equation}
where the numerator is a quadratic function of $P_J$ and the denominator is always positive:
\begin{equation}\label{}
  g(P_J)=(1+\rho P_J +P_T)(1+\rho P_J)(1+bP_J+aP_T)(1+bP_J)>0
\end{equation}
and also
\begin{equation}\label{}
  c_2=\rho b(b-\rho a)
\end{equation}
\begin{equation}\label{}
  c_1=2\rho b(a-1)
\end{equation}
\begin{equation}\label{}
  c_0=ab-\rho +a P_T (b-\rho)
\end{equation}
The stationary point of $P_J$ at which $\frac{\partial S_{A,B,x,y}}{\partial P_J}=0$ has obviously two possibilities: $\frac{c_1\pm\sqrt{c_1^2 +4c_0c_2}}{2c_2}$.

In region $\mathcal{R}_1$ where $b-\rho a>0$ and $a<1$, we have $c_2>0$ and $c_1<0$. But $c_0$ can be positive, zero or negative. If $c_0>0$, we see that $\frac{\partial S_{A,B,x,y}}{\partial P_J}$  is positive for small $P_J$ and negative for large $P_J$, and hence $S_{A,B,x,y}$ must be maximized by
\begin{equation}\label{eq:optimal_PJ2}
  P_{J,opt,x,y} = \frac{c_1+\sqrt{c_1^2 +4c_0c_2}}{2c_2}
  =\gamma+\sqrt{\gamma^2+\beta}>0
\end{equation}
But if $c_0\leq 0$, then $\frac{\partial S_{A,B,x,y}}{\partial P_J}<0$ for all $P_J\geq 0$, and hence $S_{A,B,x,y}$ is maximized by $P_J=0$.

In region $\mathcal{R}_2$ where $b-\rho a>0$ and $a\geq 1$, we have $c_2>0$, $c_1\geq 0$ and $c_0>\frac{b}{a}[a^2-1+aP_T(a-1)]\geq 0$. In this case, we see that $\frac{\partial S_{A,B,x,y}}{\partial P_J}$ is positive for small $P_J$ and negative for large $P_J$, and hence  $S_{A,B,x,y}$ is maximized by the same $P_J$ as shown in \eqref{eq:optimal_PJ2}.

In region $\mathcal{R}_3$ where $b-\rho a\leq 0$ and $a< 1$, we have $c_2\leq 0$, $c_1<0$ and $c_0\leq \frac{b}{a}[a^2-1+aP_T(a-1)]<0$, and hence $S_{A,B,x,y}$ is maximized by $P_J=0$.
\end{IEEEproof}
Recall that for region $\mathcal{R}_4$, $S_{A,B,x,y}=0$. So, only if $b-\rho a>0$ (i.e., for $(x,y)\in \mathcal{R}_1\cup\mathcal{R}_2$), $P_J$ needs to be positive. Unless mentioned otherwise, we will assume $b-\rho a>0$.

Obviously, with $b-\rho a>0$, we have $\gamma>0$ iff $a>1$. And $\beta>0$ if (not only if) $a>1$.
Also as $\rho\rightarrow 0$, $\gamma$ becomes negligible compared to $\beta$, and hence $P_{J,opt,x,y} = \sqrt{\beta}=\sqrt{\frac{a(1+P_T)}{\rho b}}$. For Figs. \ref{secrecy_capacity_single_constant_01} and \ref{secrecy_capacity_single_constant_001}, we have used this asymptotical form of $P_{J,opt,0,0}$ with large $P_T$, i.e., $P_J= \sqrt{\frac{P_T}{\rho }}$.

If we let $P_J=P_{J,opt,x,y}$, one can verify that
\begin{equation}\label{}
  \lambda_{x,y} = \left \{ \begin{array}{cc}
                       \sqrt{\frac{b}{\rho a}}, & P_T\rightarrow 0 \mbox{ and } \rho\rightarrow 0\\
                       \frac{b}{\rho a}, & P_T\rightarrow \infty
                     \end{array}
  \right .
\end{equation}

The previous results of $P_{J,opt,x,y}$ depend on $(x,y)$, which are not directly useful. We need to know the worst case of $P_{J,opt,x,y}$.

\begin{Prop}\label{Prop_of_optimal_PJ2}
Subject to $(x,y)\in \mathcal{R}_2$, i.e., $b-\rho a>0$ and $a\geq 1$:
\begin{enumerate}
  \item As $a$ increases (Eve moves towards Alice), both $\gamma$ and $\beta$ increase and hence $P_{J,opt,x,y}$ increases.
  \item As $b$ decreases (Eve moves away from Bob), both $\gamma$ and $\beta$ (although not as obvious) increases and hence $P_{J,opt,x,y}$ increases.
  \item Subject to $d_A\geq \Delta$, $\arg\max_{x,y}P_{J,opt,x,y}=(x^*,y^*)$. In other words, when Eve is at $(x^*,y^*)$, not only the secrecy is minimum but also the optimal required jamming power is maximum.
     \end{enumerate}
     \end{Prop}

     \begin{IEEEproof}
     Part 1 is obvious from \eqref{eq:optimal_PJ}. Part 2 for $\gamma$ is also obvious. To prove the property of $\beta$ in part 2,
     one can verify that
      \begin{equation}\label{}
    \frac{\partial \beta}{\partial b}=\frac{-a(1+P_T)b^2+2(1+aP_T)\rho b -a(1+aP_T)\rho^2}{\rho b^2 (b-\rho a)^2}
  \end{equation}
  where the numerator, denoted by $N(b)$, is upper bounded as follows:
  \begin{eqnarray}\label{}
    &&N(b)\leq  \max_b N(b) = N(b)|_{b=\frac{(1+aP_T)\rho}{a(1+P_T)}}\nonumber\\
     &=& -\frac{\rho^2 (1+aP_T)[(a^2-1)+a(a-1)P_T]}{a(1+P_T)}\leq 0
  \end{eqnarray}
  where the equality in the last inequality holds only for $a=1$.
   The proof of part 3 is as follows.
  For a fixed $a=\frac{1}{d_A^\alpha}$, $\max_{b=1/d_B^\alpha} P_{J,opt,x,y}$ is achieved by $b=\frac{1}{(d_A+1)^\alpha}$ which is the minimum of $b$ subject to $a=\frac{1}{d_A^\alpha}$. Furthermore, one can verify that subject to $a=\frac{1}{d_A^\alpha}$ and $b=\frac{1}{(d_A+1)^\alpha}$, both  $\gamma$ and $\beta$ increase as $d_A$ decreases. Hence, subject to $d_A\geq \Delta$, $\max_{x,y}P_{J,opt,x,y}$ is achieved by $(x^*,y^*)$.
  \end{IEEEproof}

An important implication of Property \ref{Prop_of_optimal_PJ2} is that if we know that Eves can only exist outside the disk $d_A<\Delta$, then the worst case in terms of both the secrecy $S_{A,B,x,y}$ and the optimal jamming power $P_{J,opt,x,y}$ is when Eve is at $(x^*,y^*)=(-\Delta-0.5,0)$, and the overall secrecy $S_{A,B}$ is optimized if we choose $P_J=P_{J,opt,x^*,y^*}$. Namely, the optimal value of $S_{A,B}$ is given by $S_{A,B,x^*,y^*}$ with $P_J=P_{J,opt,x^*,y^*}$.

\subsection{With small-scale fading}
We now consider small-scale fading. In this case,
 the secrecy capacity $S_{A,B,x,y}$ is still given by \eqref{eq:SABxy} with \eqref{eq:CAB} and \eqref{eq:CAE}. But the SNRs in \eqref{eq:SNRAB} and \eqref{eq:SNRAE} need to be revised as follows:
\begin{equation}\label{}
  SNR_{A,B}=\frac{\tilde A P_T}{1+\rho \tilde B P_J}
\end{equation}
\begin{equation}\label{}
  SNR_{A,E}=\frac{a \tilde C  P_T}{1+b \tilde D P_J}
\end{equation}
where $\tilde A,\tilde B,\tilde C,\tilde D$ are small scale fading factors. We can assume that Alice and Bob know $\tilde A$ and $\tilde B$ but not $\tilde C$ and $\tilde D$. In principle, Alice and Bob can make decisions based on the knowledge of $\tilde A$ and $\tilde B$.
Since the large-scale fading is already taken care of, we can assume
 that
$\tilde A$, $\tilde B$, $\tilde C$ and $\tilde D$ are i.i.d. (independent and identically distributed) and each has the exponential pdf (probability density function) $e^{-u}$ with $u\geq 0$. Note that the amplitudes of all channel gains are proportional to the square-roots of these factors and hence are Rayleigh-distributed. Due to the random nature of $\tilde C$ and $\tilde D$ in particular, $S_{A,B,x,y}$ is now random. We will be interested in probabilities of zero secrecy and/or their upper bounds, which should be made small for good security.

\begin{figure}
  \centering
  \includegraphics[width=4.5cm]{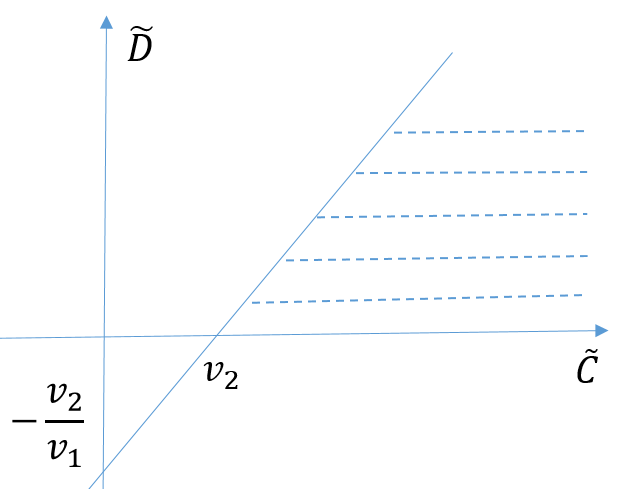}\\
  \caption{The shaded region is defined by \eqref{eq:CDregion}.}\label{shade1}
\end{figure}

We know that the secrecy $S_{A,B,x,y}$ is zero if and only if $SNR_{A,B}\leq SNR_{A,E}$ or equivalently,
\begin{equation}\label{eq:CDregion}
 \tilde C-v_1 \tilde D-v_2\geq 0
\end{equation}
with
\begin{equation}\label{eq:v1}
  v_1 =
   \frac{b\tilde AP_J}{a(1+\rho \tilde BP_J)}
\end{equation}
\begin{equation}\label{eq:v2}
  v_2 = \frac{\tilde A}{a(1+\rho \tilde BP_J)}
\end{equation}
Notice that if $P_J$ increases, $v_1$ increases and $v_2$ decreases. The region defined by \eqref{eq:CDregion} is illustrated by the shaded area in Fig. \ref{shade1}.

\begin{Prop} \label{Prop_of_zero_secrecy}
\begin{enumerate}
  \item The conditional probability of zero secrecy, conditional upon $\tilde A$ and $\tilde B$, is
\begin{eqnarray}\label{eq:PSABxy0}
  &&\mathcal{P}_{\{S_{A,B,x,y}=0|\tilde A,\tilde B\}}=\int_0^\infty dv \int_{v_1 v + v_2}^\infty e^{-u-v}du\nonumber\\
  &=&\frac{e^{-v_2}}{1+v_1}<\frac{1}{1+v_1}
\end{eqnarray}
\item Subject to $d_A\geq \Delta$,  $\arg\max_{x,y}\mathcal{P}_{\{S_{A,B,x,y}=0|\tilde A,\tilde B\}}=(x^*,y^*)$, i.e., the worst location of Eve is the same as shown before.
\item If Eve is arbitrarily close to Alice, the probability of zero secrecy is one, i.e., $\mathcal{P}_{\{S_{A,B,-0.5,0}=0|\tilde A,\tilde B\}}=\mathcal{P}_{\{S_{A,B,-0.5,0}=0\}}=1$.
  \item In general, the unconditional probability of zero secrecy  is:
  \begin{equation}\label{eq:upper_bound}
  \mathcal{P}_{\{S_{A,B,x,y}=0\}}=\mathcal{E}\{\mathcal{P}_{\{S_{A,B,x,y}=0|\tilde A,\tilde B\}}\}<\mathcal{E}\left \{\frac{1}{1+v_1}\right \}
\end{equation}
where the upper bound is a decreasing function\footnote{Due to $v_2$ in \eqref{eq:PSABxy0}, the monotonic property of the exact $\mathcal{P}_{\{S_{A,B,x,y}=0\}}$ in terms of $P_J$ has so far no proof.} of $P_J$ and is maximized when Eve is at $(x^*,y^*)$ subject to $d_A\geq \Delta$.
\end{enumerate}
\end{Prop}
\begin{IEEEproof}
For Part 1, we see that $\mathcal{P}_{\{S_{A,B,x,y}=0|\tilde A,\tilde B\}}$ is equivalent to $\mathcal{P}_{\{\tilde C-v_1 \tilde D-v_2\geq 0\}}$ which equals the integral of the joint pdf of $\tilde C$ and $\tilde D$ over the shaded region shown in Fig. \ref{shade1}.
 For part 2, note that $\mathcal{P}_{\{S_{A,B,x,y}=0|\tilde A,\tilde B\}}$ decreases as $v_1$ and/or $v_2$ increase, or equivalently, as $a$ decreases and/or $b$ increases. So, for a fixed $a=\frac{1}{d_A^\alpha}$, $\arg\max_b\mathcal{P}_{\{S_{A,B,x,y}=0|\tilde A,\tilde B\}}=\arg\min_b v_1=\frac{1}{(d_A+1)^\alpha}$. Then, subject to $a=\frac{1}{d_A^\alpha}$, $b=\frac{1}{(d_A+1)^\alpha}$ and $d_A\geq \Delta$, $\arg\max_{d_A}\mathcal{P}_{\{S_{A,B}=0|\tilde A,\tilde B\}}=\arg\min_{d_A} v_1=\arg\min_{d_A} v_2=\Delta$.  Part 3 follows by using $a=\infty$ and $b=1$ (i.e., $v_1=0$ and $v_2=0$) in \eqref{eq:PSABxy0}. Part 4 is obvious.
\end{IEEEproof}

Although the optimal jamming power in terms of the upper bound \eqref{eq:upper_bound} on the unconditional probability of zero secrecy $\mathcal{P}_{\{S_{A,B,x,y}=0\}}$ is infinite, it is not yet clear whether the optimal jamming power in terms of the conditional probability of zero secrecy $\mathcal{P}_{\{S_{A,B,x,y}=0|\tilde A,\tilde B\}}$ is also infinite. Keep in mind that the choice of $P_J$ can be based on $\tilde A$ and $\tilde B$ which are known to Alice and Bob.

One can verify that
\begin{eqnarray}\label{eq:derivative}
  &&\frac{\partial \mathcal{P}_{\{S_{A,B,x,y}=0|\tilde A,\tilde B\}}}{\partial P_J} \nonumber\\
  &=& \frac{-\tilde A e^{-v_2}(a_1 P_J+a_0)}{(a(1+\rho \tilde B P_J) + b\tilde AP_J)^2(1+\rho \tilde B P_J)}
\end{eqnarray}
with $a_1=\rho \tilde B[a(b-\rho \tilde B)-b\tilde A]$ and $a_0=a(b-\rho \tilde B)$. We see that each of $a_0$ and $a_1$ can be either positive or negative. If $a_0>0$ and $a_1>0$, then $\mathcal{P}_{\{S_{A,B,x,y}=0|\tilde A,\tilde B\}}$ is always a decreasing function of $P_J$ and hence the optimal value $P_{J,opt,x,y}$ of $P_J$ is infinite. If $a_0>0$ and $a_1<0$, then there is a finite optimal power, i.e., $P_{J,opt,x,y}=\frac{a_0}{-a_1}$. If $a_0<0$, which also implies $a_1<0$, then $\mathcal{P}_{\{S_{A,B,x,y}=0|\tilde A,\tilde B\}}$ is always an increasing function of $P_J$, for which $P_{J,opt,x,y}=0$.


We see that depending on the realizations of $\tilde A$ and $\tilde B$, the optimal jamming power could be zero, finite or infinite.
However, one can verify that the condition where $a_0>0$ and $a_1>0$ is equivalent to $\tilde B<\frac{b}{\rho}$ and $\tilde A< a(1-\frac{\rho}{b}\tilde B)$, which is also equivalent to
$\tilde A<a$ and $\tilde B<\frac{b}{\rho}(1-\frac{\tilde A}{a})$. This observation leads to:

\begin{Prop}\label{Prob_of_monotone}
\begin{enumerate}
  \item A lower bound on the probability of $\mathcal{P}_{\{S_{A,B,x,y}=0|\tilde A,\tilde B\}}$ being a decreasing function of $P_J$ is:
\begin{eqnarray}\label{}
  &&Prob\left \{\Tilde A<a \mbox{ and } \tilde B<\frac{b}{\rho}\left (1-\frac{\tilde A}{a}\right ) \right \}
  \nonumber\\
  &=&\int_0^a du \int_0^{\frac{b}{\rho}(1-\frac{u}{a})} e^{-u-v}dv\nonumber\\
  &=&1-\frac{b}{b-\rho a}e^{-a}+\frac{\rho a}{b-\rho a} e^{-\frac{b}{\rho}}\doteq \underline{\mathcal{P}}
\end{eqnarray}
\item Let $\frac{b}{\rho}=\eta a$ with $\eta>1$. Then,
\begin{equation}\label{}
  \underline{\mathcal{P}}=1-\frac{1}{\eta-1}[\eta e^{-a}-e^{-\eta a}]
  =\left \{ \begin{array}{cc}
              1, & a=\infty \\
              1-e^{-a}, & \eta=\infty
            \end{array}
  \right .
\end{equation}
and $\underline{\mathcal{P}}$ increases rapidly to one as $a$ increases for any $\eta>1$.
\end{enumerate}
\end{Prop}
\begin{IEEEproof}
The proof is straightforward.
\end{IEEEproof}

For example, if $\Delta=0.1$, $\alpha=2$ and $\eta=1.01$ (i.e., $\rho\approx -21$dB), then at $(x^*,y^*)$, $\underline{\mathcal{P}}=1-2.3\times10^{-42}$. This example suggests although in general the optimal jamming power may be finite depending on $\tilde A$ and $\tilde B$, almost surely the (worst-case) conditional probability of zero secrecy is a decreasing function of $P_J$ for which the optimal jamming power is infinite\footnote{In practice, this should be translated into a maximum allowed jamming power.}.

\begin{figure}
  \centering
  \includegraphics[width=5cm]{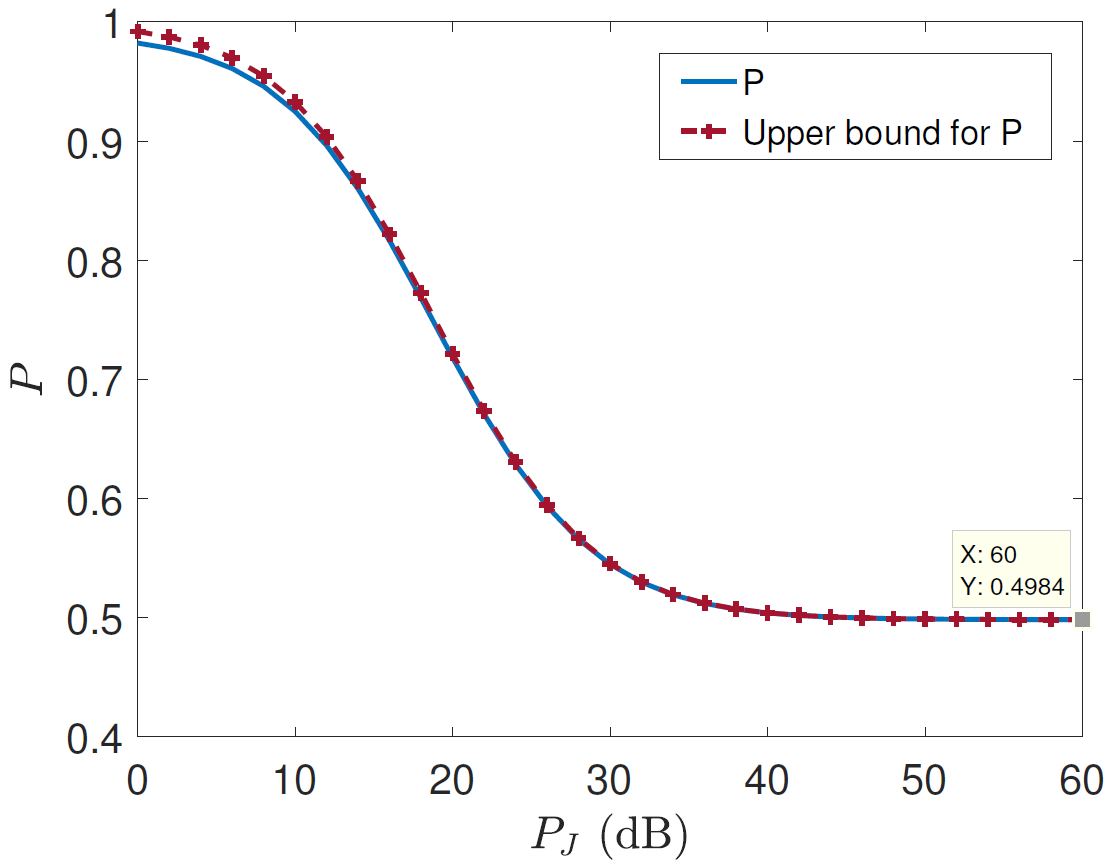}\\
  \caption{$\mathcal{P}_{\{S_{A,B,x^*,y^*}=0\}}$ and its upper bound given by \eqref{eq:upper_bound} with $\Delta=0.1$, $\alpha=2$ and $\eta=1.01$ ($\rho=21$dB).}\label{prob_of_zero_secrecy_upper_bound}
\end{figure}

      Shown in Fig. \ref{prob_of_zero_secrecy_upper_bound} are the unconditional probability of zero secrecy $\mathcal{P}_{\{S_{A,B,x^*,y^*}=0\}}$ and its upper bound given by \eqref{eq:upper_bound} with $\Delta=0.1$, $\alpha=2$ and $\eta=1.01$. The gap of the bound is small (intuitively because of the small $v_2$ in $e^{-v_2}$ in \eqref{eq:PSABxy0}). We see that $\mathcal{P}_{\{S_{A,B,x^*,y^*}=0\}}$ is about $0.5$ for $P_J\geq 40$dB. If we use 10 transmissions of 10 keys under independent small-scale fading, we could have the unconditional probability of zero secrecy guaranteed to be less than $2^{-10}$ everywhere subject to $d_A\geq 0.1$. Note that zero secrecy of the 10 transmissions happens iff each of the 10 transmissions has zero secrecy. In other words, if any of the transmissions has a positive secrecy, the overall secrecy is positive\footnote{In practice, one would need a bit of extra margin of a positive secrecy so that the security is reliable subject to unknown errors.}. To achieve independent small-scale fading between transmissions in scattering-rich environments, Alice and Bob can simply move their locations by a short distance in the order of half-wavelength after each transmission.

\begin{figure}
  \centering
  \includegraphics[width=5cm]{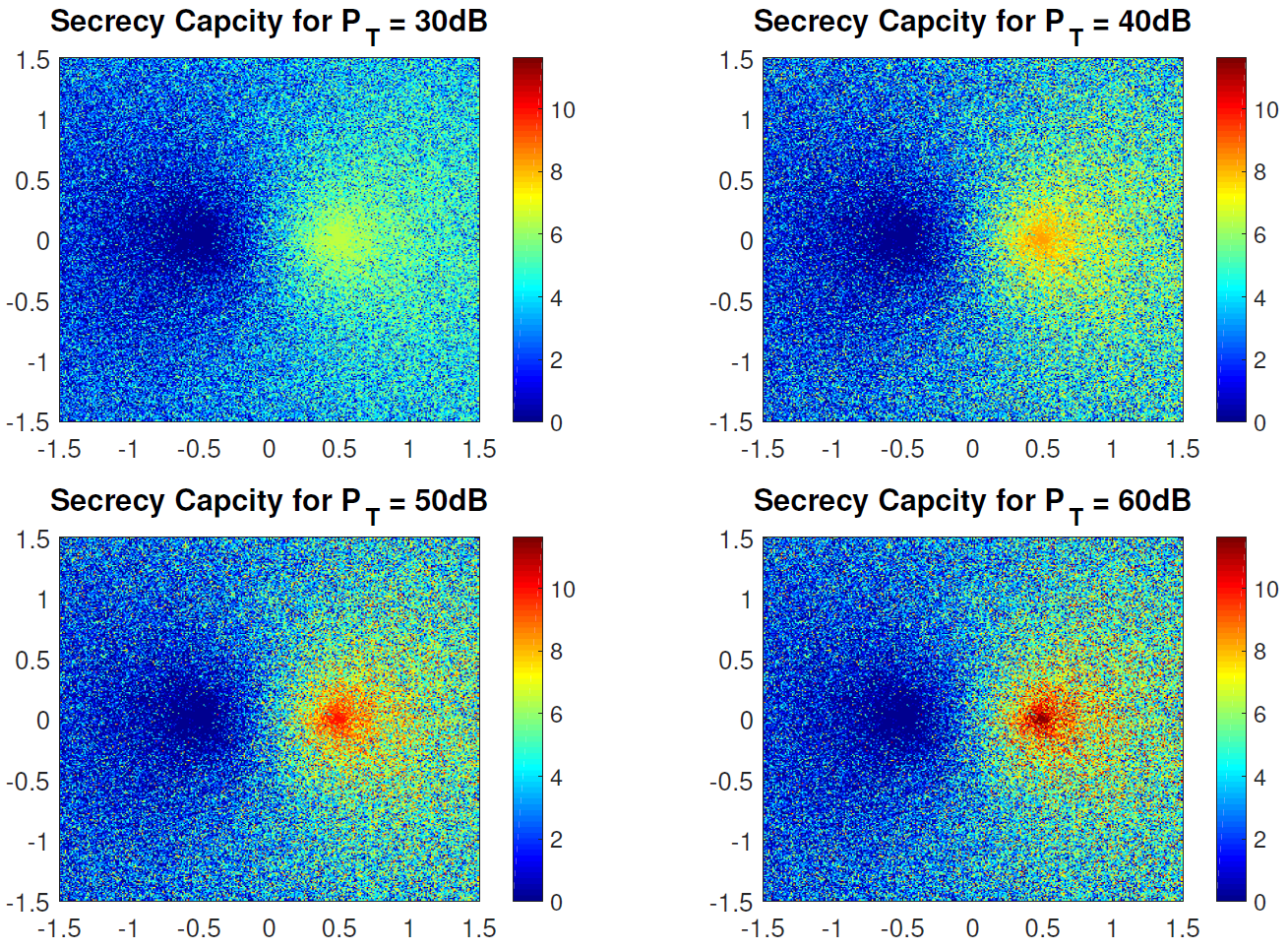}\\
  \caption{With small-scale fading and subject to $\tilde A=\tilde B=1$, $S_{A,B,x,y}$ versus $(x,y)$ with $P_J=\sqrt{\frac{P_T}{\rho}}$, $\rho=0.1$ and $\alpha=2$.
  }\label{secrecy_capacity_single_fading_01}
\end{figure}

\begin{figure}
  \centering
  \includegraphics[width=5cm]{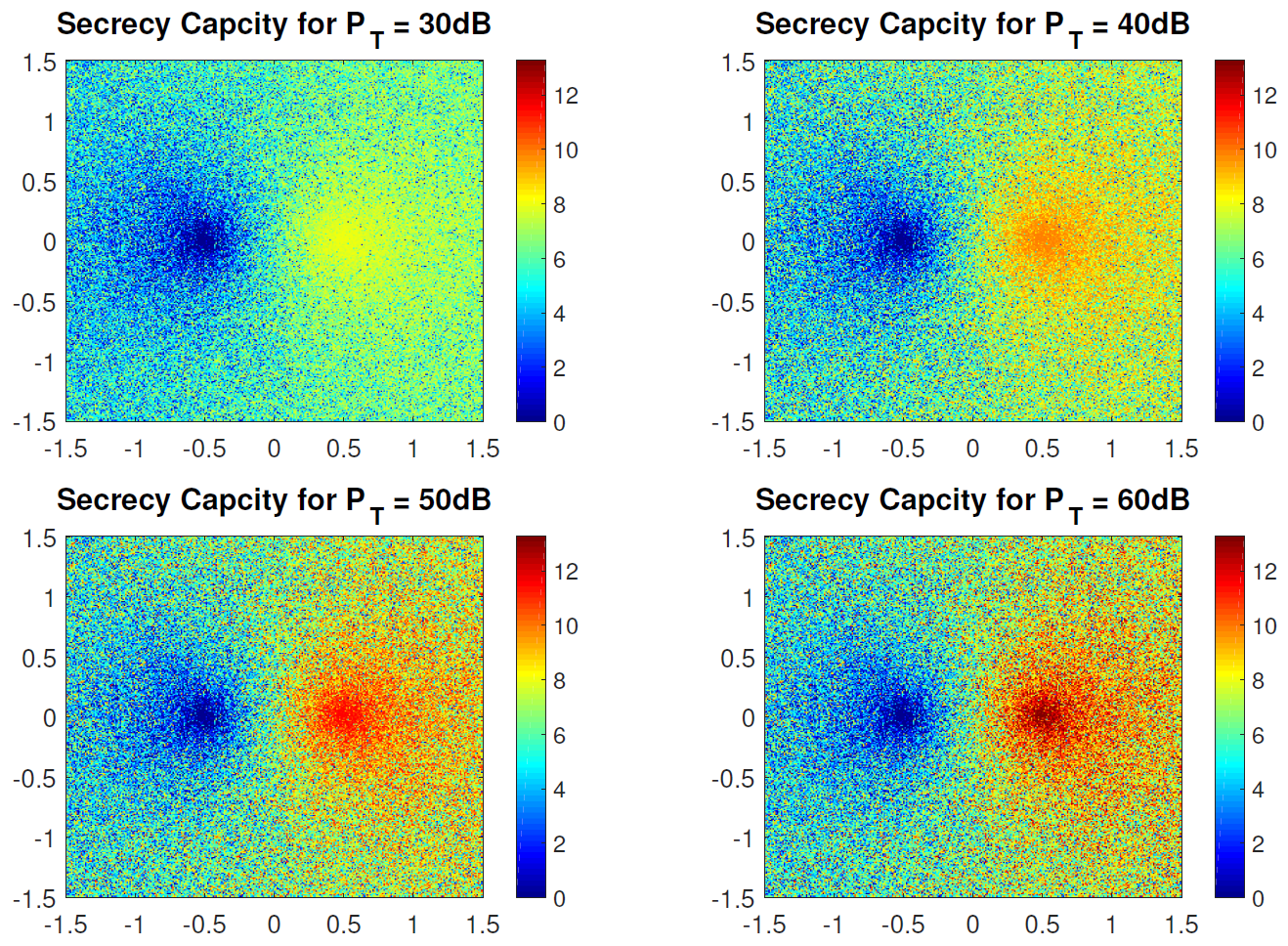}\\
  \caption{With small-scale fading and subject to $\tilde A=\tilde B=1$, $S_{A,B,x,y}$ versus $(x,y)$ with $P_J=\sqrt{\frac{P_T}{\rho}}$, $\rho=0.01$ and $\alpha=2$.
   }\label{secrecy_capacity_single_fading_001}
\end{figure}

Shown in Fig. \ref{secrecy_capacity_single_fading_001} and \ref{secrecy_capacity_single_fading_01} are examples of $S_{A,B,x,y}$ versus $(x,y)$ with small-scale fading subject to $\tilde A=\tilde B=1$. In these figures, the step size in each of $x$ and $y$ directions is 0.01. For each sample of $(x,y)$, an independent realization of $\tilde C$ and $\tilde D$ is used. We see that due to small-scale fading, even in region $\mathcal{R}_4$ the conditional probability of zero secrecy is not zero (unless $a=\infty$ or $d_A=0$).

The conditional probability of zero secrecy $\mathcal{P}_{\{S_{A,B,x,y}=0|\tilde A,\tilde B\}}$ in \eqref{eq:PSABxy0} can be used by Alice and Bob for opportunistic transmission of secret keys. In scattering-rich environment, $\tilde A$ and $\tilde B$ can change significantly after a small amount (in the order of half-wavelength) of location change of Bob. (Note that a location change of Alice would affect $\tilde A$, $\tilde C$ and $\tilde D$ significantly but likely do little for $\tilde B$. This is because the self-interference channel of Bob is mainly affected by objects around Bob.)
In particular, for mobile ad hoc network (MANET), Alice and Bob could both move around until  $\mathcal{P}_{\{S_{A,B,x,y}=0|\tilde A,\tilde B\}}$ is small enough. Based on \eqref{eq:v1}, \eqref{eq:v2} and \eqref{eq:PSABxy0}, we see that $\mathcal{P}_{\{S_{A,B,x,y}=0|\tilde A,\tilde B\}}$ is minimized by the largest $\tilde A$ and the smallest $\tilde B$.

For opportunistic transmission of secret keys, it is useful to consider
the cumulative distribution function (CDF) of  $\mathcal{P}_{\{S_{A,B,x,y}=0|\tilde A,\tilde B\}}$ as well as its dependence on $P_J$. A closed-form CDF of the exact $\mathcal{P}_{\{S_{A,B,x,y}=0|\tilde A,\tilde B\}}$ seems intractable to find. But using the upper bound in \eqref{eq:PSABxy0}, one can verify the following lower bound on the CDF of $\mathcal{P}_{\{S_{A,B,x,y}=0|\tilde A,\tilde B\}}$:
\begin{eqnarray}\label{eq:CDF_Lowerbound}
  \mathcal{F}_{\{A,B,x,y\}}(p)&\doteq & Prob\{\mathcal{P}_{\{S_{A,B,x,y}
  =0|\tilde A,\tilde B\}}\leq p\}\nonumber\\
  &>&Prob\left \{\frac{1}{1+v_1}\leq p\right \}\nonumber\\
  &=&e^{-\frac{a(1-p)}{bP_Jp}}\frac{bp}{bp+a\rho (1-p)}
\end{eqnarray}
where $0< p\leq 1$. A simple dependence of the lower bound of $\mathcal{F}_{\{A,B,x,y\}}(p)$ on $P_J$ is very clear in \eqref{eq:CDF_Lowerbound}.
One can also verify that if $P_J=\infty$,
\begin{equation}\label{eq:CDFPJlarge}
  \mathcal{F}_{\{A,B,x,y\}}(p)
  =\frac{bp}{bp+a\rho (1-p)}
\end{equation}
which means that the inequality in \eqref{eq:CDF_Lowerbound} becomes equality under $P_J=\infty$.

\begin{figure}
  \centering
  \includegraphics[width=5cm]{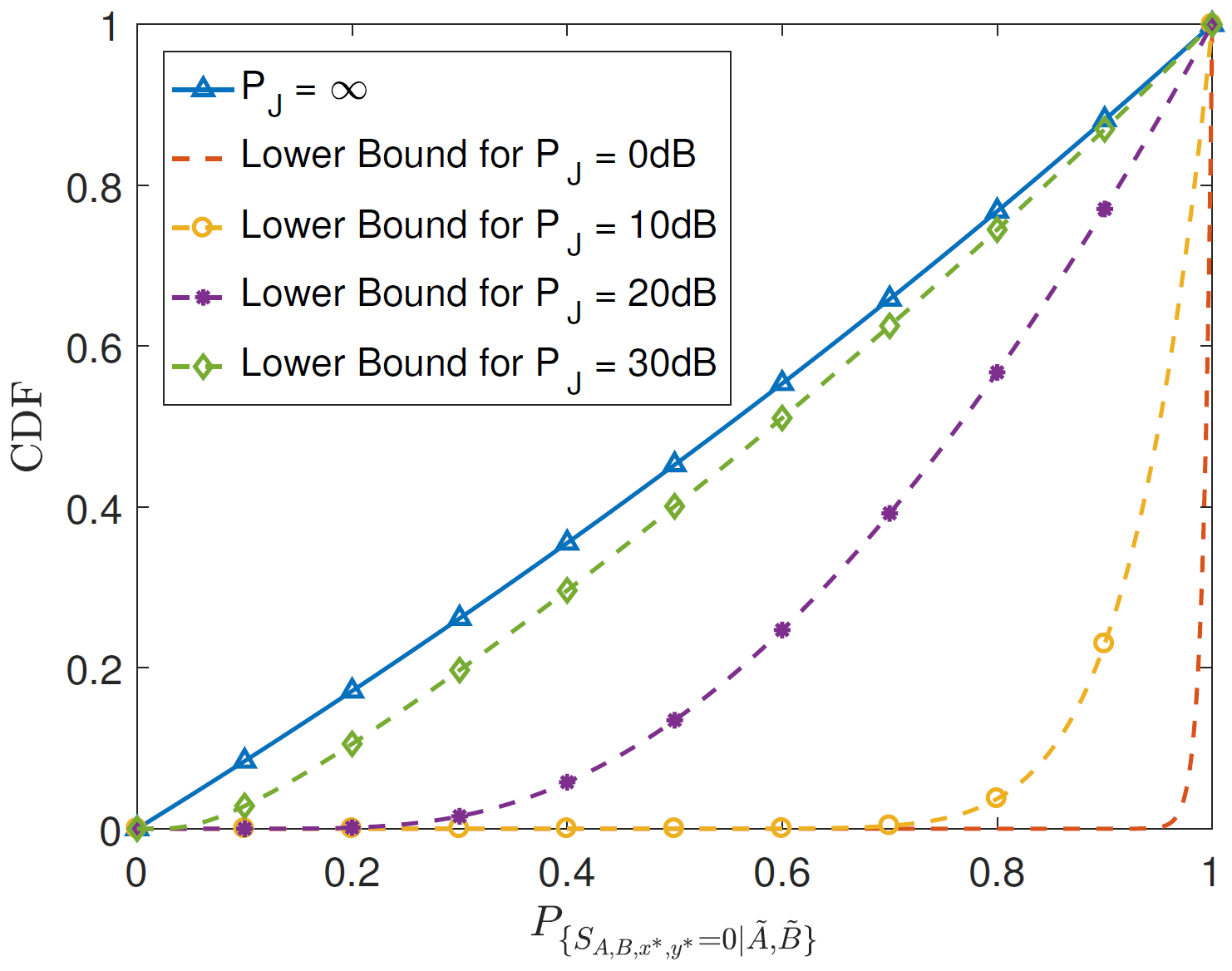}\\
  \caption{Lower bound on CDF of $\mathcal{P}_{\{S_{A,B,x^*,y^*}=0|\tilde A,\tilde B\}}$ with $\rho=0.01$ and $(x^*,y^*)=(-\Delta-0.5,0)=(-0.6,0)$. }\label{CDF_single_001}
\end{figure}

Shown in Fig. \ref{CDF_single_001} is the CDF lower bound given by \eqref{eq:CDF_Lowerbound}. For example, we see that at $P_J=30dB$, there is at least a $10\%$ chance (with respect to the random $\tilde A$ and $\tilde B$ that are known to Alice and Bob) that $\mathcal{P}_{\{S_{A,B,x^*,y^*}=0|\tilde A,\tilde B\}}<20\%$ (with respect to the random $\tilde C$ and $\tilde D$ that are unknown to Alice and Bob).  In other words, for any given $0<p<1$, the larger is the CDF lower bound, the more likely can Alice and Bob encounter such $\tilde A$ and $\tilde B$ that the conditional probability of zero secrecy is less than $p$.

\section{Without collusion among eavesdroppers}\label{sec:no_collusion}
Now we consider the case where there is no collusion among Eves. Namely, all Eves act as isolated individuals. To reveal the fundamentals, we continue to assume the simple system discussed so far. But in this section, we also assume that Alice and Bob are both full-duplex radios and of equal characteristics in terms of $P_T$, $P_J$ and $\rho$.

To take the advantage of no collusion among Eves, we consider a dual-phase transmission scheme. In phase 1, Alice first sends a random key $K_1$ to Bob while Bob receives the key and also jams Eves in full-duplex mode. In phase 2, Bob sends another random key $K_2$ to Alice as Alice receives the key and also jam Eves in full-duplex mode. Following the dual-phase scheme, each of the two nodes has the same pair of keys $(K_1,K_2)$ as their secret information.

The above scheme was inspired by a scheme in \cite{Katabi2011} where a so-called iJam was proposed for conventional half-duplex radios. In phase 1 of iJam, Alice sends a random key $K_1$ over $N$ subcarriers to Bob while Bob jams Eves over a random subset $\mathcal{S}_1$ of $\frac{N}{2}$ subcarriers and receives samples over the other subset $\mathcal{S}_2$ of $\frac{N}{2}$ subcarriers. Then, Alice repeats the same process while Bob jams Eves over $\mathcal{S}_2$ and receives samples over $\mathcal{S}_1$. Combining the un-jammed samples received during the two transmissions in phase 1, Bob recovers $K_1$. In phase 2 of iJam, the roles of Alice and Bob are reversed, and after two repeated transmissions from Bob, Alice receives another key $K_2$ from Bob.

To compare our scheme with iJam conveniently, we will call our scheme fJam.  The main differences between fJam and iJam are:
\begin{enumerate}
  \item For the same pair of keys, iJam could require up to twice as much spectral resource as fJam requires.
  \item The header of each transmitted packet by iJam is not jammed and is completely transparent  to Eves while fJam jams the entire packet of each transmission. It is important to note that the header of a packet also carries important information such as pilots for channel estimation.
  \item A random $\mathcal{S}_1$ in iJam has a finite number $C^{\frac{N}{2}}_N=\frac{N!}{(\frac{N}{2}!)^2}<2^N$ of possibilities.  Under iJam, Eve is in theory able to recover both keys through exhaust search although the computational complexity is increased by $(C^{\frac{N}{2}}_N)^2$ times when compared to no jamming. In other words, the secrecy capacity of iJam is actually zero.
  But under fJam, the secrecy capacity can be easily made positive, and no Eve is even in theory able to recover both keys correctly with infinite computational power.
\end{enumerate}

\subsection{Without small-scale fading}

As long as the time taken between the two phases is small enough, we can assume that the location and channel response of any Eve is unchanged. The secrecy capacity of fJam against Eve at unknown location $(x,y)$ is therefore:
       \begin{equation}\label{eq:Sxy}
         S_{x,y}=\frac{1}{2}(S_{A,B,x,y}+S_{B,A,x,y})
       \end{equation}
       with $S_{A,B,x,y}=(C_{A,B}-C_{A,E})^+$, $S_{B,A,x,y}=(C_{B,A}-C_{B,E})^+$, $C_{A,B}=\log_2(1+SNR_{A,B})$, $C_{B,A}=\log_2(1+SNR_{B,A})$, $C_{A,E}=\log_2(1+SNR_{A,E})$, $C_{B,E}=\log_2(1+SNR_{B,E})$ and
       \begin{equation}\label{}
         SNR_{A,B}=SNR_{B,A}=\frac{P_T}{1+\rho P_J}\doteq SNR
       \end{equation}
         \begin{equation}\label{}
         SNR_{A,E}=\frac{aP_T}{1+bP_J}
       \end{equation}
       \begin{equation}\label{}
         SNR_{B,E}=\frac{bP_T}{1+aP_J}
       \end{equation}
The overall secrecy capacity of fJam against all Eves should be $S=\min_{x,y}S_{x,y}$, which could also be subject to some constraint on $(x,y)$ if there is any prior knowledge of the locations of Eves. Clearly, in order to understand $S$, it is sufficient to understand $S_{x,y}$ as a function of $(x,y)$. In particular, we are interested in the worst cases of $S_{x,y}$.

We see that $S_{x,y}$ is symmetric between $a$ and $b$ since it is an average of $S_{A,B,x,y}$ and $S_{B,A,x,y}$. Recall Property \ref{Prop1a}.  There are four unique regions of $S_{A,B,x,y}$, i.e., $\mathcal{R}_i$ with $i=1,2,3,4$, which can be rewritten as $\mathcal{R}_i=\mathcal{R}_i(a,b)$ to stress its dependence on $a$ and $b$. Therefore, for $S_{B,A,x,y}$, there must be corresponding four regions which we can denote by $\mathcal{\bar R}_i$ with $i=1,2,3,4$, and $\mathcal{\bar R}_i=\mathcal{R}_i(b,a)$. Similarly, we will write $\mathcal{R}_{\rho}=\mathcal{R}_{\rho}(a,b)$ and $\mathcal{\bar R}_{\rho}=\mathcal{R}_{\rho}(b,a)$. Also, $\gamma=\gamma (a,b)$ and $\bar {\gamma} = \gamma(b,a)$. Therefore, we have:

\begin{Prop}\label{basic_property_dual}
\begin{enumerate}
 \item If $P_J=0$, then $S_{x,y}>0$ iff $(x,y)\notin \mathcal{R}_{d_A\leq 1}\cap\mathcal{R}_{d_B\leq 1}$.
  \item ``There is $P_J\geq 0$ such that $S_{x,y}>0$'' iff $(x,y) \notin \mathcal{R}_4\cap \mathcal{\bar R}_4$.
  \item ``For arbitrary $(x,y)$, there is $P_J\geq 0$ such that $S_{x,y}>0$'' iff $\rho<1$.
\end{enumerate}
\end{Prop}
\begin{IEEEproof}
The proof is easy based on Property \ref{Prop1a}.
\end{IEEEproof}

\begin{figure}
  \centering
  \includegraphics[width=4.5cm]{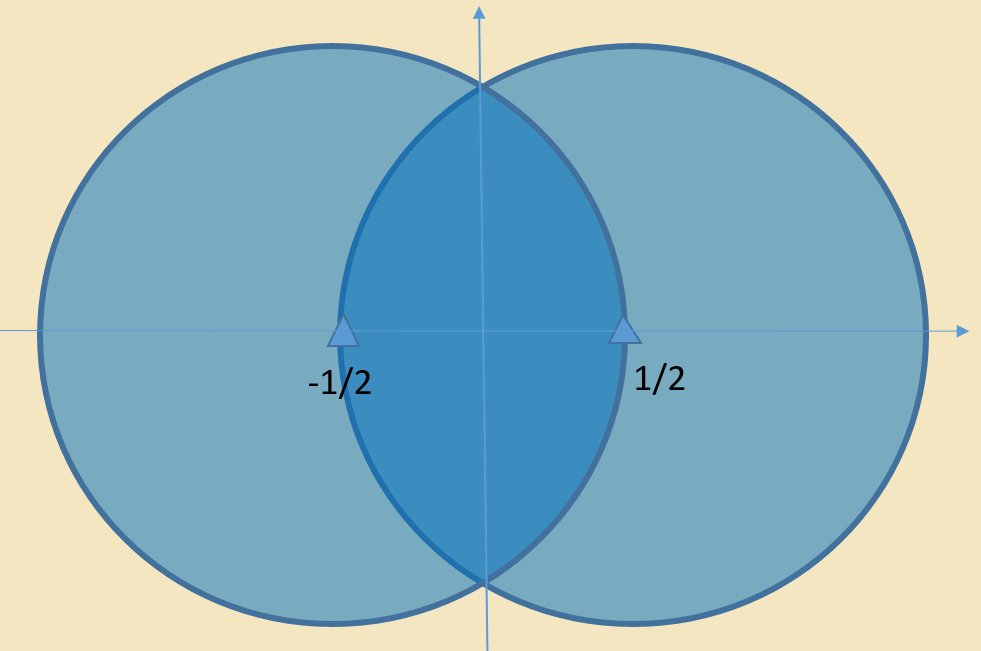}\\
  \caption{With $P_J=0$, $S_{x,y}>0$ iff $(x,y)$ is not in the dark blue region (i.e., the ``almond-shaped'' region).}\label{almond}
\end{figure}

\begin{figure}
  \centering
  \includegraphics[width=4.5cm]{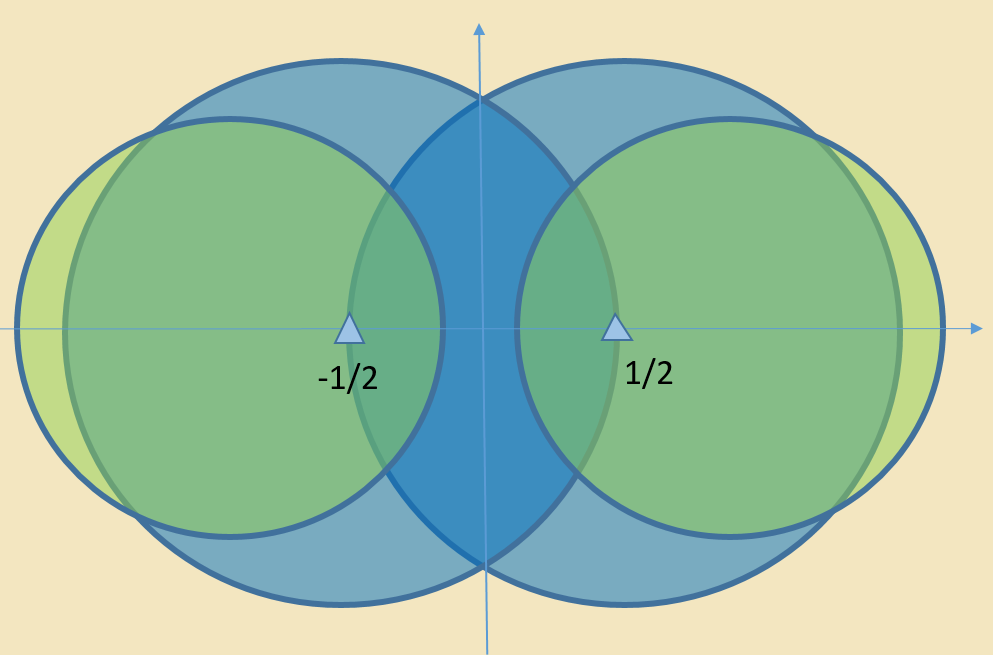}\\
  \caption{``There is $P_J\geq 0$ such that $S_{x,y}>0$'' iff $(x,y)$ is not in the dark blue region which vanishes iff $\rho<1$.}\label{almond_with_partial_removal}
\end{figure}

Part 1 of Property \ref{basic_property_dual} is illustrated in Fig. \ref{almond}. Part 2  of Property \ref{basic_property_dual} is illustrated in Fig. \ref{almond_with_partial_removal} with $\rho>1$. Part 3 says that in order to have a positive secrecy against Eve at any location we must have $\rho<1$. We will assume $\rho<1$.

The general landscape of $S_{x,y}$ is a function of $\rho$, $P_T$ and $P_J$, and is more complicated than that of $S_{A,B,x,y}$. In particularly, unlike $S_{A,B,x,y}$, $S_{x,y}$ does not in general increase or decrease monotonically with respect to either $a$ or $b$. But an important and tractable region of $S_{x,y}$ is along the $x$-axis and the $y$-axis, which will be focused on next.

\begin{Prop}\label{Prop_of_Sxy}
Assume  $(x,y)\in \mathcal{R}_\rho\cap\mathcal{\bar R}_\rho$ and $P_J>\max\{\gamma,\bar {\gamma}\}$. Then, $S_{0,0}$ is the minimum of $S_{x,y}$ along the $y$-axis, and $S_{0,y}$ increases as $|y|$ increases.
\end{Prop}
\begin{IEEEproof}
It is easy to verify that under the stated condition,
\begin{equation}\label{eq:STxy}
  S_{x,y}=\log_2(1+SNR)-\frac{1}{2}\log_2 T_{x,y}
\end{equation}
with
\begin{eqnarray}\label{eq:Txy}
  T_{x,y}&=&(1+SNR_{A,E})(1+SNR_{B,E})\nonumber\\
  &=&\frac{(1+bP_J+aP_T)(1+aP_J+bP_T)}{(1+bP_J)(1+aP_J)}
\end{eqnarray}
Let $x=0$. Then, $a=b=\frac{1}{(0.25+y^2)^{\alpha/2}}$ and  $T_{0,y}=\left ( 1+\frac{aP_T}{1+aP_T}\right )^2$  which decreases as $|y|$ increases and has its maximum at $y=0$. Hence, $S_{0,y}$ increases as $|y|$ increases and has its minimum at $y=0$.
\end{IEEEproof}

But along the $x$-axis, $S_{x,y}$ is no longer as monotonic as it is along the $y$-axis. The pattern of $S_{x,0}$ is highly sensitive to the choice of $P_T$ and $P_J$:
\begin{Prop}\label{Prop_Sx0}
Assume $\rho<1$ and $P_J>\frac{1-2^{-\alpha}}{1-\rho}$. Then, $S_{0,0}$ is a local maximum along the $x$-axis if
\begin{equation}\label{eq:Px}
  P_J>\frac{-1+\sqrt{1+2^{\alpha+1}P_T}}{2^{\alpha+1}}
\end{equation}
If the inequality is reversed, $S_{0,0}$ is a local minimum along the $x$-axis.
\end{Prop}
\begin{IEEEproof}
The stated condition complies with that of Property \ref{Prop_of_Sxy} and hence \eqref{eq:STxy} holds. One can verify that for $y=0$ and a small $x\neq 0$,
\begin{eqnarray}\label{eq:Tx0}
  &&T_{x,0}=T_{0,0}+\nonumber\\
  &&\frac{\alpha^2 2^{2(\alpha+1)}}{(1+2^\alpha P_J)^2}\left [ \left (1+\frac{2^\alpha P_T}{1+2^\alpha P_J} \right )^2 P_J^2 - (P_J-P_T)^2\right ]x^2\nonumber\\
\end{eqnarray}
If $P_J\geq P_T$, \eqref{eq:Tx0} obviously implies that $T_{x,0}>T_{0,0}$ and hence $T_{0,0}$ ($S_{0,0}$) is a local minimum (maximum) along the $x$-axis. But more generally, $S_{0,0}$ is a local maximum along the $x$-axis if the coefficient of $x^2$ in \eqref{eq:Tx0} is positive. One can verify that this condition is equivalent to $2^{\alpha+1}P_J^2+2P_J-P_T>0$ which is also equivalent to \eqref{eq:Px}. The converse is obvious.
\end{IEEEproof}

We now look deeper into $S_{x,0}$.
For convenience, we let $d=d_A$, $a=\frac{1}{d^\alpha}$ and $b=\frac{1}{(1-d)^\alpha}$. Then $\log_2T_{x,0}=\log_2 T_{d-0.5,0}$ which is a function of $d$ with $d=x+0.5$. Assuming the condition of Property \ref{Prop_of_Sxy}, one can verify that\footnote{For compact expressions, we assume $\alpha$ is even.}
\begin{equation}\label{eq:derivative of Sxy}
  \frac{\partial \log_2S_{x,0}}{\partial x} =-\frac{1}{2}\frac{\partial \log_2 T_{x,0}}{\partial x}= -\frac{\log_2e}{2}\frac{N(d)}{ D(d)}
\end{equation}
 with
\begin{eqnarray}\label{}
  D(d)&=&\left (1+\frac{P_J}{(1-d)^\alpha}+\frac{P_T}{d^\alpha}\right )\left (1+\frac{P_J}{(1-d)^\alpha}\right )\nonumber\\
  &\cdot &\left (1+\frac{P_J}{d^\alpha}+\frac{P_T}{(1-d)^\alpha}\right )\left (1+\frac{P_J}{d^\alpha}\right)>0
\end{eqnarray}
\small
\begin{eqnarray}
  &&N(d) = - \alpha P_JP_T^2 \frac{d^{\alpha -1} -(1-d)^{\alpha -1}}{d^{2\alpha}(1-d)^{2\alpha}}\nonumber \\
  &+&\alpha P_T\frac{d^{\alpha +1}-(1-d)^{\alpha +1}}{d^{\alpha +1}(1-d)^{\alpha +1}}
   + 2\alpha P_JP_T\frac{d^{2\alpha +1}-(1-d)^{2\alpha +1}}{d^{2\alpha +1}(1-d)^{2\alpha +1}}
   \nonumber \\
   &+& \alpha P_J^2P_T \left ( 2\frac{d^\alpha -(1-d)^\alpha }{d^{2\alpha +1} (1-d)^{2\alpha +1}}+\frac{d^{3\alpha+1}-(1-d)^{3\alpha+1}}{d^{3\alpha+1}(1-d)^{3\alpha+1}}\right )
   \nonumber\\
   &+&\alpha P_J^3P_T \frac{d^{2\alpha}-(1-d)^{2\alpha}}{d^{3\alpha +1}(1-d)^{3\alpha +1}}+  \alpha P_T^2 \frac{2d-1}{d^{\alpha +1}(1-d)^{\alpha+1}}
\end{eqnarray}
\normalsize
Due to the symmetry $S_{x,0}=S_{-x,0}$, we only need to consider $x\geq 0$ (i.e., $d\geq 0.5$).
We see that all terms in $N(d)$ are negative if $d>1$. Hence, under the condition of Property \ref{Prop_of_Sxy}, $S_{x,0}$ is an increasing function of $x$ for $x>0.5$. It is also interesting to see that if $1>d>0.5$,
all terms in $N(d)$ except the first term are positive, and the first term is dominated by some other terms if $P_J$ is large enough.  So, if $P_J$ is large enough, $S_{x,0}$ must be a decreasing function of $x$ for $0<x<0.5$, which is consistent with Property \ref{Prop_Sx0}.

Also note that all terms in $N(d)$ except the first term change their sign as $d$ increases across the value of one. Furthermore, one can verify:
\begin{Prop}
Under the condition of Property \ref{Prop_of_Sxy}, if we let $x\rightarrow 0.5$, then
\begin{equation}\label{eq:singularity}
  \frac{\partial \log_2S(x,0)}{\partial x}=-\frac{\log_2e}{2}\frac{\alpha}{0.5-x}
\end{equation}
\end{Prop}
\begin{IEEEproof}
The proof is based on \eqref{eq:derivative of Sxy}, where only the dominant term is kept.
\end{IEEEproof}

This property shows a singularity of the derivative of $S_{x,0}$ near $x=0.5$. However, one can also verify that under the condition of Property \ref{Prop_of_Sxy},
the requirement $x\rightarrow 0.5$ means that $P_J\rightarrow \infty$ and $\rho\rightarrow 0$. So, \eqref{eq:singularity} should be appreciated only under these asymptotical conditions. Clearly, for a fixed pair of $P_J$ and $\rho$, \eqref{eq:singularity} does not reflect the actual pattern of $S_{x,0}$ when $x$ is so close to $0.5$ that $(x,0)\notin\mathcal{R}_\rho \cap \mathcal{\bar R}_\rho$ or $P_J\leq \max\{\gamma,\bar \gamma\}$, which will be further discussed later.

We now compare the secrecy between the left and right sides of Bob.
At a $\Delta$-distance to the left of Bob, $a=\frac{1}{(1-\Delta)^\alpha}$ and $b=\frac{1}{\Delta^\alpha}$. Under the condition of Property \ref{Prop_of_Sxy} with $\Delta\ll 1$ and hence $P_J\gg 1$ (but $\frac{P_T}{P_J}$ can be arbitrary), one can verify that $SNR_{A,E}=\Delta^\alpha(1+\alpha\Delta)\frac{P_T}{P_J}$, $SNR_{B,E}=\Delta^{-\alpha}(1-\alpha\Delta)\frac{P_T}{P_J}$ and hence
\begin{eqnarray}\label{}
  &&T_{0.5-\Delta,0} \nonumber\\
  && =\left (1+\Delta^\alpha(1+\alpha\Delta)\frac{P_T}{P_J}\right )\left (1+\Delta^{-\alpha}(1-\alpha\Delta)\frac{P_T}{P_J}\right )\nonumber\\
  && =1+\Delta^{-\alpha}(1-\alpha\Delta)\frac{P_T}{P_J}+(1-\alpha^2\Delta^2)\frac{P_T^2}{P_J^2}
\end{eqnarray}
Similarly, at the same $\Delta$-distance but to the right of Bob, one can verify that
\begin{equation}\label{}
  T_{0.5+\Delta,0}=
  1+\Delta^{-\alpha}(1+\alpha\Delta)\frac{P_T}{P_J}+(1-\alpha^2\Delta^2)\frac{P_T^2}{P_J^2}
\end{equation}
We see that $T_{0.5+\Delta,0}-T_{0.5-\Delta,0}=2\alpha \Delta^{1-\alpha}\frac{P_T}{P_J}>0$. It follows:

\begin{Prop}
Under the condition of Property \ref{Prop_of_Sxy}, the secrecy capacity at a short distance to the right of Bob (or the left of Alice) is smaller than that at the same distance to the left of Bob (or the right of Alice).
\end{Prop}


The optimal jamming power to maximize $S_{x,y}$ is difficult to analyze in general due to the need to find the roots of a 4th-order polynomial in $P_J$. If we use the origin as the reference location, we know $S_{0,0}=S_{A,B,0,0}=S_{B,A,0,0}$ and hence $\arg\max_{P_J}S_{0,0}=\arg\max_{P_J}S_{A,B,0,0}=P_{J,opt,0,0}$ where $P_{J,opt,x,y}$ is given in \eqref{eq:optimal_PJ}.
 For small $\rho$ and large $P_T$, one can verify that
\begin{equation}\label{}
  P_{J,opt,0,0}=\sqrt{\frac{P_T}{\rho}}
\end{equation}

\begin{Prop} \label{constant_secrecy}
If $P_J=\sqrt{\frac{P_T}{\rho}}$ which is invariant to $(x,y)$, $\rho<\frac{\Delta^\alpha}{(1+\Delta)^\alpha}$, $1>d_A> \Delta$, $1>d_B> \Delta$, and $\Delta^\alpha \sqrt{\rho P_T}\gg 1$,   then
\begin{equation}\label{eq:constant_secrecy}
  S_{x,y}=\log_2\frac{1}{\rho}
\end{equation}
which is invariant to $(x,y)$.
\end{Prop}
\begin{IEEEproof}
It follows that $\frac{a}{b}\sqrt{\rho P_T}> \Delta^\alpha a \sqrt{\rho P_T}>\Delta^\alpha  \sqrt{\rho P_T}\gg 1$ and $a\sqrt{\frac{P_T}{\rho}}>a\sqrt{\rho P_T}>\sqrt{\rho P_T}>\Delta^\alpha \sqrt{\rho P_T}\gg 1$. Similarly, $\frac{b}{a}\sqrt{\rho P_T}\gg 1$, $b\sqrt{\frac{P_T}{\rho}}\gg 1$ and $\sqrt{\rho P_T}\gg 1$. Then, one can verify that $S_{A,B,x,y}=\log_2\frac{1}{\rho}-\log_2\frac{a}{b}>0$ and $S_{B,A,x,y}=\log_2\frac{1}{\rho}-\log_2\frac{b}{a}>0$, which implies \eqref{eq:constant_secrecy}.
\end{IEEEproof}

\begin{figure}
  \centering
  \includegraphics[width=5cm]{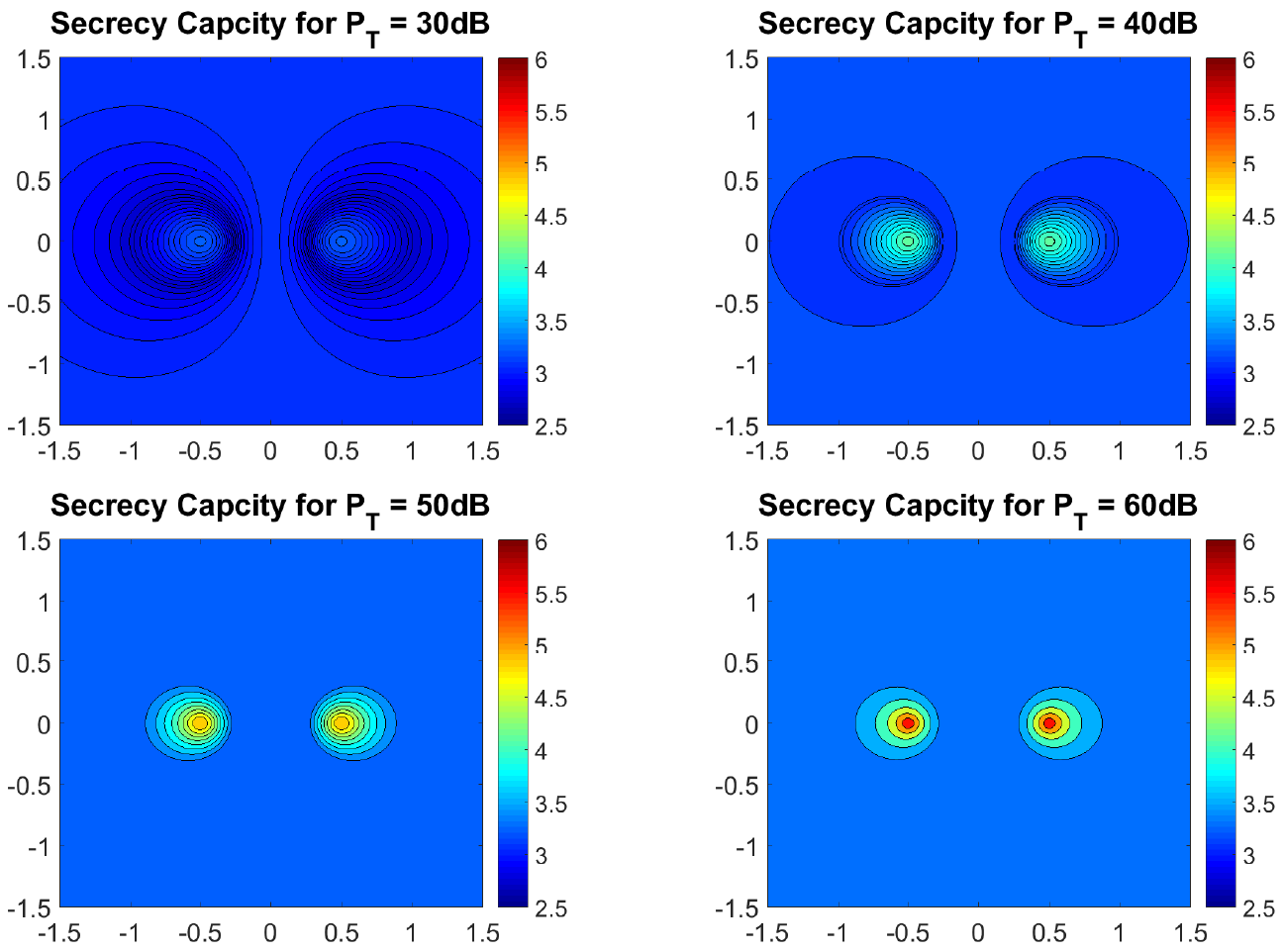}\\
  \caption{``Flat near-field'' with $P_J=\sqrt{\frac{P_T}{\rho}}$, $\alpha=2$, $\rho=0.1$ (i.e., $\log_2\frac{1}{\rho}=3.3$).
   }\label{secrecy_capacity_dual_constant_01}
\end{figure}

\begin{figure}
  \centering
  \includegraphics[width=5cm]{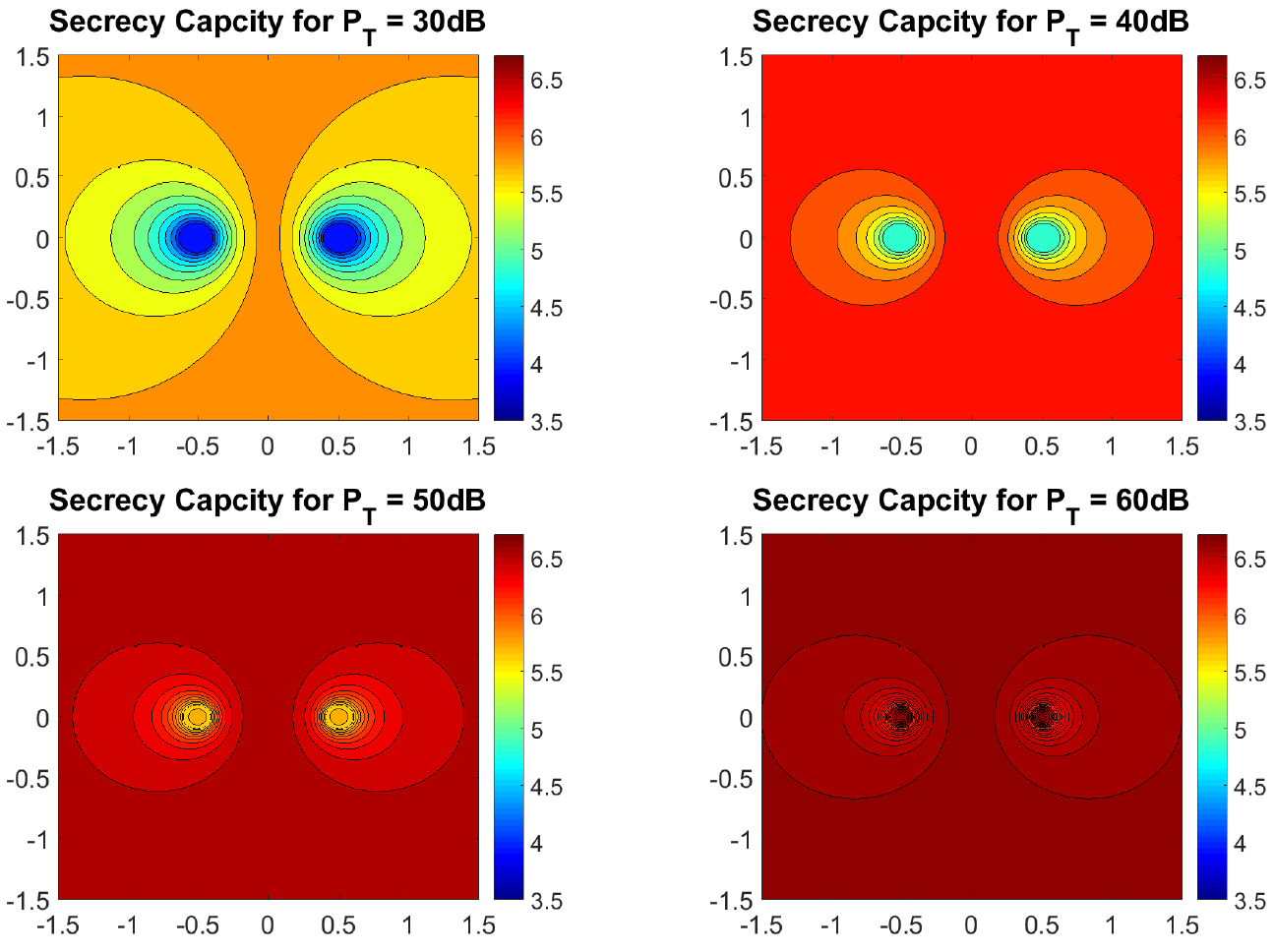}\\
  \caption{``Flat near-field'' with $P_J=\sqrt{\frac{P_T}{\rho}}$, $\alpha=2$, $\rho=0.01$ (i.e., $\log_2\frac{1}{\rho}=6.6$).
   }\label{secrecy_capacity_dual_constant_001}
\end{figure}

This property shows that under large $P_T$, small $\rho$ and $P_J=\sqrt{\frac{P_T}{\rho}}$, there is a constant secrecy region in a ``near-field''.
But if $\sqrt{\rho P_T}\gg 1$ and Eve is far way from Alice and Bob, then one can verify that $S_{x,y}=\frac{1}{2}\log_2\frac{P_T}{\rho}$, which is a constant secrecy in a ``far-field'' and as expected much larger than that in the near-field. Illustrated in Figs. \ref{secrecy_capacity_dual_constant_01} and \ref{secrecy_capacity_dual_constant_001} is the constant near-field secrecy capacity. We see that at $P_T=60$dB, \eqref{eq:constant_secrecy} is a good approximation of the near-field secrecy for both $\rho=0.1$ and $\rho=0.01$.


We now consider the case where $(x,y)$ is so close to $(\pm0.5,0)$ that $(x,y)\notin\mathcal{R}_\rho \cap \mathcal{\bar R}_\rho$ or $P_J\leq \max\{\gamma,\bar \gamma\}$ (i.e., one of the conditions of Property of \ref{Prop_of_Sxy} is violated).

\begin{Prop}
Assume $\rho<\frac{1}{2^\alpha}$ so that $\mathcal{R}_3$ and $\mathcal{\bar R}_3$ do not exist, and $\mathcal{R}_4$ and $\mathcal{\bar R}_4$ are the complements of $\mathcal{R}_\rho$ and $\mathcal{\bar R}_\rho$ respectively. Recall $\mathcal{R}_{P_J}=\{(x,y)\in\mathcal{R}_\rho|P_J\leq \gamma\}$ and define $\mathcal{\bar R}_{P_J}=\{(x,y)\in\mathcal{\bar R}_\rho|P_J\leq \bar \gamma\}$. Then,
\begin{enumerate}
  \item For $(x,y)\in \mathcal{R}_4\cup \mathcal{R}_{P_J}$: $S_{x,y}=\frac{1}{2}S_{B,A,x,y}$ and, if $P_J>0$, $\max_{x,y}S_{x,y}=S_{-0.5,0}=\frac{1}{2}\log_2 ( 1+SNR )$.
  \item For $(x,y)\in \mathcal{\bar R}_4\cup \mathcal{\bar R}_{P_J}$: $S_{x,y}=\frac{1}{2}S_{A,B,x,y}$ and, if $P_J>0$, $\max_{x,y}S_{x,y}=S_{0.5,0}=\frac{1}{2}\log_2 ( 1+SNR )$.
\end{enumerate}
\end{Prop}
\begin{IEEEproof}
First recall Property \ref{Prop_of_ring}. Then consider that for $P_J>0$,  $SNR_{A,E}$ approaches zero as $(x,y)$ approaches $(0.5,0)$, and $SNR_{B,E}$ approaches zero as $(x,y)$ approaches $(-0.5,0)$.
\end{IEEEproof}
This property says that with $P_J>0$, $S_{x,y}$ actually peaks locally at the locations of Alice and Bob (although, under a large enough $P_J$, $S_{x,y}$ decreases as the location of Eve moves along the $x$-axis from the origin towards Alice or Bob initially when the condition of Property \ref{Prop_of_Sxy} holds).  This property is clearly visible in Fig. \ref{secrecy_capacity_dual_constant_01}, but is difficult to see in \ref{secrecy_capacity_dual_constant_001}. The latter is because the regions $\mathcal{R}_4\cup \mathcal{R}_{P_J}$ (around Alice) and $\mathcal{\bar R}_4\cup \mathcal{\bar R}_{P_J}$ (around Bob) with $\rho=0.01$ are too small within which the variation of $S_{x,y}$ is small.

\subsection{With small-scale fading}
\begin{figure}
  \centering
  \includegraphics[width=4.5cm]{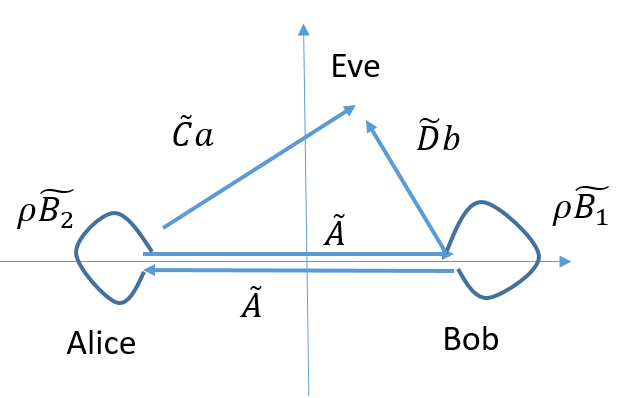}\\
  \caption{Illustration of small-scale fading factors for fJam.}\label{fading_fJam}
\end{figure}

With small-scale fading, the secrecy capacity is still given by \eqref{eq:Sxy} but with
\begin{equation}\label{eq:SNRs_fading1}
         SNR_{A,B}=\frac{\tilde A P_T}{1+\rho \tilde B_1 P_J}
       \end{equation}
       \begin{equation}\label{eq:SNRs_fading2}
         SNR_{B,A}=\frac{\tilde A P_T}{1+\rho \tilde B_2 P_J}
       \end{equation}
         \begin{equation}\label{eq:SNRs_fading3}
         SNR_{A,E}=\frac{a\tilde C P_T}{1+b \tilde D P_J}
       \end{equation}
       \begin{equation}\label{eq:SNRs_fading4}
         SNR_{B,E}=\frac{b\tilde D P_T}{1+a \tilde C P_J}
       \end{equation}
where the small-scale fading factors are illustrated in Fig. \ref{fading_fJam}. In particular, we assume that $\tilde A$ is the same for either transmission from Alice to Bob or transmission from Bob to Alice (justified by the reciprocal property of electro-magnetics). We treat $\tilde B_1$ and $\tilde B_2$ as different small-scale fading factors because the self-interference at Alice and that at Bob are affected differently by surrounding scattering objects due to the location difference of Alice and Bob. As in the previous section, we will assume that all small-scale fading factors are i.i.d. exponentially distributed with unit mean. Keep in mind that Alice and Bob can make decisions based on the knowledge of $\tilde A$, $\tilde B_1$ and $\tilde B_2$.

\begin{figure}
  \centering
  \includegraphics[width=4.5cm]{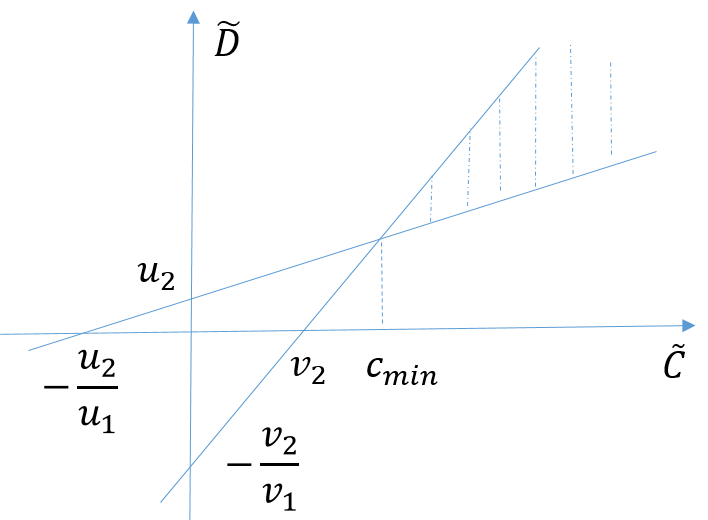}\\
  \caption{Illustration of the integrated (shaded) area in \eqref{eq:2D_integral}.}\label{2D_integral}
\end{figure}

It follows that $S_{x,y}=0$ iff $C_{A,B}\leq C_{A,E}$ and $C_{B,A}\leq C_{B,E}$ or equivalently iff
\begin{equation}\label{eq:double_linear_inequality}
\left \{ \begin{array}{c}
           \tilde C-v_1 \tilde D-v_2\geq 0,\\
           \tilde D-u_1 \tilde C-u_2\geq 0.
         \end{array}
\right .
\end{equation}
where $v_1=\frac{b\tilde AP_J}{a(1+\rho \tilde B_1P_J)}$, $v_2=\frac{\tilde A}{a(1+\rho \tilde B_1 P_J)}$, $u_1=\frac{a\tilde AP_J}{b(1+\rho \tilde B_2 P_J)}$ and $u_2=\frac{\tilde A}{b(1+\rho \tilde B_2 P_J)}$. The pair of inequalities in \eqref{eq:double_linear_inequality} corresponds to the shaded area in Fig. \ref{2D_integral}.
Then, the probability of zero secrecy conditional upon $\tilde A$, $\tilde B_1$ and $\tilde B_2$ is
\begin{eqnarray}\label{eq:2D_integral}
  &&\mathcal{P}_{\{S_{x,y}=0|\tilde A,\tilde B_1,\tilde B_2\}}
  =\int_{c_{min}}^\infty dx \int_{u_1x+u_2}^\frac{x-v_2}{v_1}  e^{-x-y}dy\nonumber\\
  &&=\frac{1}{1+u_1}e^{-u_2-(1+u_1)c_{min}}-
  \frac{1}{1+{\frac{1}{v_1}}}e^{\frac{v_2}{v_1}-(1+\frac{1}{v_1})c_{min}}\nonumber\\
\end{eqnarray}
where
\begin{equation}\label{eq:cmin}
  c_{min}=\left \{ \begin{array}{cc}
                     \frac{v_2+v_1u_2}{1-v_1u_1}, & v_1u_1<1 \\
                     \infty & \mbox{otherwise}
                   \end{array}
  \right .
\end{equation}

One can further verify from \eqref{2D_integral} and \eqref{eq:cmin} that
\begin{equation}\label{eq:PSxy0}
  \mathcal{P}_{\{S_{x,y}=0|\tilde A,\tilde B_1,\tilde B_2\}}=\left \{ \begin{array}{cc}
                                                                    Ke^{-E}, &  w_1>0\\
                                                                    0 & \mbox{otherwise}
                                                                  \end{array}
  \right .
\end{equation}
where $K=\frac{w_1}{w_2}$, $E=\frac{w_3}{w_1}$ and
\begin{equation}\label{eq:w1}
  w_1=ab[(1+\rho \tilde B_1 P_J)(1+\rho \tilde B_2 P_J)-\tilde A^2 P_J^2]
\end{equation}
\begin{equation}\label{}
  w_2=[a\tilde A P_J +b(1+\rho \tilde B_2 P_J)][b\tilde A P_J+a(1+\rho \tilde B_1 P_J)]
\end{equation}
\begin{equation}\label{}
  w_3=\tilde A[a(1+\rho\tilde B_1 P_J)+b(1+\rho\tilde B_2 P_J)+b\tilde A P_J+a\tilde A P_J]
\end{equation}

\begin{Prop}
With zero jamming power, the unconditional probability of zero secrecy anywhere is upper bounded by that at origin, i.e.,
\begin{equation}\label{}
   \mathcal{P}_{\{S_{x,y}=0|P_J=0\}}\leq \mathcal{P}_{\{S_{0,0}=0|P_J=0\}}=\frac{1}{1+ 0.5^{\alpha-1}}
\end{equation}
Also, for an Eve arbitrarily close to Alice or Bob, we have $\mathcal{P}_{\{S_{\pm 0.5,0}=0|P_J=0\}}=0.5$.
\end{Prop}
\begin{IEEEproof}
If $P_J=0$, then $K=1$ and $E=\tilde A\frac{a+b}{ab}$. Hence $\mathcal{P}_{\{S_{x,y}=0|P_J=0\}}=\mathcal{E}\{\mathcal{P}_{\{S_{x,y}=0|\tilde A,\tilde B_1,\tilde B_2,P_J=0\}}\}=\mathcal{E}\{e^{-\tilde A \frac{a+b}{ab}}\}=\frac{1}{1+\frac{a+b}{ab}}$. Given $a$, $\frac{a+b}{ab}$ is a decreasing function of $b$. So, for any given $d_A$, $\max_{x,y}\mathcal{P}_{\{S_{x,y}=0|P_J=0\}}$ is achieved by the smallest $d_B$. So, $\max_{x,y}\mathcal{P}_{\{S_{x,y}=0|P_J=0\}}$ is achieved only if $y=0$. Now consider  $y=0$ and $\mathcal{P}_{\{S_{x,0}=0|P_J=0\}}=\frac{1}{1+|x+0.5|^\alpha+|x-0.5|^\alpha}$ which is symmetric of $x$ and has its peak at $x=0$. And obviously $\mathcal{P}_{\{S_{\pm 0.5,0}=0|P_J=0\}}=0.5$.
\end{IEEEproof}

If $P_J=\infty$,
\begin{equation}\label{}
  K=\frac{ab(\rho^2 \tilde B_1 \tilde B_2 -\tilde A^2)}{(a\tilde A+\rho b\tilde B_2)
  (b\tilde A+\rho a\tilde B_1)}
\end{equation}
\begin{equation}\label{}
  E=\frac{\tilde A (\rho a\tilde B_1+\rho b\tilde B_2+b\tilde A+a\tilde A)}{ab(\rho^2 \tilde B_1 \tilde B_2 -\tilde A^2)P_J}\rightarrow 0
\end{equation}
Hence,
\begin{eqnarray}\label{}
  &&\mathcal{P}_{\{S_{x,y}=0|P_J=\infty\}}=\mathcal{E}\{K\}\nonumber\\
  &=&
  \int_0^\infty\int_0^\infty\int_0^{\rho\sqrt{uv}}Ke^{-u-v-w}dudvdw
\end{eqnarray}
where the random variables $\tilde A, \tilde B_1, \tilde B_2$ in $K$ should be replaced by the dummy variables $w,u,v$ respectively (this rule will be used again later).
Note that $\arg\max_{x,y}K$ is generally a function of $\tilde A$, $\tilde B_1$ and $\tilde B_2$. Unlike the case of $P_J=0$, it seems intractable to find an analytical form of $\arg\max_{x,y}\mathcal{P}_{\{S_{x,y}=0|P_J=\infty\}}$.

For the general case of $P_J>0$, searching for the worst location of Eve in terms of $\mathcal{P}_{\{S_{x,y}=0|\tilde A,\tilde B_1,\tilde B_2\}}$ or $\mathcal{P}_{\{S_{x,y}=0\}}$ is also hard. But the following special case should be of interest:

\begin{Prop}\label{Prop_of_Eve_at_Alice_and_Bob}
If $P_J>0$ and Eve is arbitrarily close to Alice or Bob, then the probability of zero secrecy is zero, i.e., $\mathcal{P}_{\{S_{\pm 0.5,0}=0|P_J>0\}}=0$.
\end{Prop}
\begin{IEEEproof}
Assume any $P_J>0$. At $(x,y)=(0.5,0)$, we have $a=1$ and $b=\infty$, and hence  $v_1=\infty$, $v_2=\frac{\tilde A}{1+\rho \tilde B P_J}$, $u_1=0$ and $u_2=0$. In this case, $v_1u_1$ is not defined and  the previous derivation does not apply. To see the corresponding result, we consider the necessary condition for zero secrecy:   $C-v_1 D-v_2\geq 0$, which now holds with zero probability (since $v_1=\infty$). Hence, $\mathcal{P}_{\{S_{\pm 0.5,0}=0|\tilde A,\tilde B_1,\tilde B_2,P_J>0\}}=0$ and hence $\mathcal{P}_{\{S_{\pm 0.5,0}=0|P_J>0\}}=0$.
\end{IEEEproof}

Property \ref{Prop_of_Eve_at_Alice_and_Bob} is in contrast to the case of colluding Eves where the probability of zero secrecy at Alice is always one. See Part 3 in Property \ref{Prop_of_zero_secrecy}.

\begin{Prop}\label{Prop_of_dynamicPJ}
Iff $\tilde A^2 > \rho^2 \tilde B_1 \tilde B_2$: then there is a $P_J^*>0$ such that iff $P_J\geq P_J^*$, $\mathcal{P}_{\{S_{x,y}=0|\tilde A,\tilde B_1,\tilde B_2\}}=0$, where
\begin{equation}\label{}
  P_J^*=\frac{\rho (\tilde B_1+\tilde B_2)+\sqrt{\rho^2(\tilde B_1+\tilde B_2)^2+4(\tilde A^2-\rho^2\tilde B_1 \tilde B2)}}
  {2(\tilde A^2-\rho^2 \tilde B_1 \tilde B_2)}
\end{equation}
Namely, iff $\tilde A^2 > \rho^2 \tilde B_1 \tilde B_2$, Alice and Bob can choose a $P_J$ to make the probability of zero secrecy equal to zero anywhere.
\end{Prop}
\begin{IEEEproof}
The proof follows from \eqref{eq:PSxy0} and \eqref{eq:w1}.
\end{IEEEproof}

Property \ref{Prop_of_dynamicPJ} can be used for dynamic control of jamming power. A semi-dynamic control is:
Alice and Bob choose $P_J=P_J^*$ if $\tilde A^2 > \rho^2 \tilde B_1 \tilde B_2$, or a constant $P_J$ if $\tilde A^2 \leq \rho^2 \tilde B_1 \tilde B_2$.

\begin{Prop}\label{Prop_of_Prob_Sxy0_dynamic_PJ}
With the above semi-dynamic control of jamming power, we have
\begin{equation}\label{}
  \mathcal{P}_{\{S_{x,y}=0\}}=
  \int_0^\infty\int_0^\infty\int_0^{\rho \sqrt{uv}} Ke^{-E}e^{-u-v-w}dudvdw<\mathcal{P}_1
\end{equation}
with $\mathcal{P}_1=Prob\{\tilde A^2 \leq \rho^2 \tilde B_1 \tilde B_2\}$, i.e.,
\begin{equation}
 \mathcal{P}_1
  =\int_0^\infty \int_0^\infty e^{-u-v}[1-e^{-\rho\sqrt{uv}}]dudv
  <\frac{\pi \rho}{4}
\end{equation}
where the last inequality becomes tight as $\rho$ becomes small.
\end{Prop}
\begin{IEEEproof}
$\mathcal{P}_{\{S_{x,y}=0\}}$ is the expectation of $\mathcal{P}_{\{S_{x,y}=0|\tilde A,\tilde B_1,\tilde B_2\}}$ in \eqref{eq:PSxy0} subject to $\tilde A^2 \leq \rho^2 \tilde B_1 \tilde B_2$. Using $Ke^{-E}<1$, we have
\begin{eqnarray}\label{eq:P1}
 \mathcal{P}_{\{S_{x,y}=0\}}&<&\mathcal{P}_1\doteq \int_0^\infty  \int_0^\infty  \int_0^{\rho\sqrt{uv}} e^{-u-v-w}dudvdw\nonumber\\
  &=&\int_0^\infty \int_0^\infty e^{-u-v}[1-e^{-\rho\sqrt{uv}}]dudv\nonumber\\
  &<& \int_0^\infty \int_0^\infty e^{-u-v}\rho\sqrt{uv}dudv=\frac{\pi \rho}{4}
\end{eqnarray}
where the second inequality is due to $1-e^{-z}<z$ for any $z>0$, and the last equality follows from the fact $\int_0^\infty e^{-u}\sqrt{u}du=\frac{\sqrt{\pi}}{2}$. This fact can be proved by using the change of variable $u=\frac{v^2}{2}$, which leads to $\int_0^\infty e^{-u}\sqrt{u}du=\frac{\sqrt{\pi}}{2}\sigma^2$ where $\sigma^2= \frac{2}{\sqrt{2\pi}}\int_0^\infty v^2 e^{-\frac{v^2}{2}}dv=1$ which is known as the variance of a Gaussian random variable with zero mean and unit variance.
\end{IEEEproof}

\begin{Prop}\label{Prop_of_Prob_Sxy0_constant_PJ}
If $P_J$ is a constant regardless of $\tilde A^2 > \rho^2 \tilde B_1 \tilde B_2$, then
\begin{equation}\label{}
  \mathcal{P}_{\{S_{x,y}=0\}}=
  \int_0^\infty\int_0^\infty\int_0^{w_0} Ke^{-E}e^{-u-v-w}dudvdw<\mathcal{P}_2
\end{equation}
where $w_0=\sqrt{\rho^2 uv+\frac{1+\rho (u+v)P_J}{P_J^2}}$ and
  \begin{eqnarray}\label{}
  \mathcal{P}_2
  &=&\int_0^\infty \int_0^\infty e^{-u-v}\left [1-e^{-w_0}\right ]dudv >\mathcal{P}_1
\end{eqnarray}
and
for $P_J=\infty$, $\mathcal{P}_1=\mathcal{P}_2$.
\end{Prop}
\begin{IEEEproof}
Similar to the proof of Property \ref{Prop_of_Prob_Sxy0_dynamic_PJ}.
\end{IEEEproof}

\begin{figure}
  \centering
  \includegraphics[width=5cm]{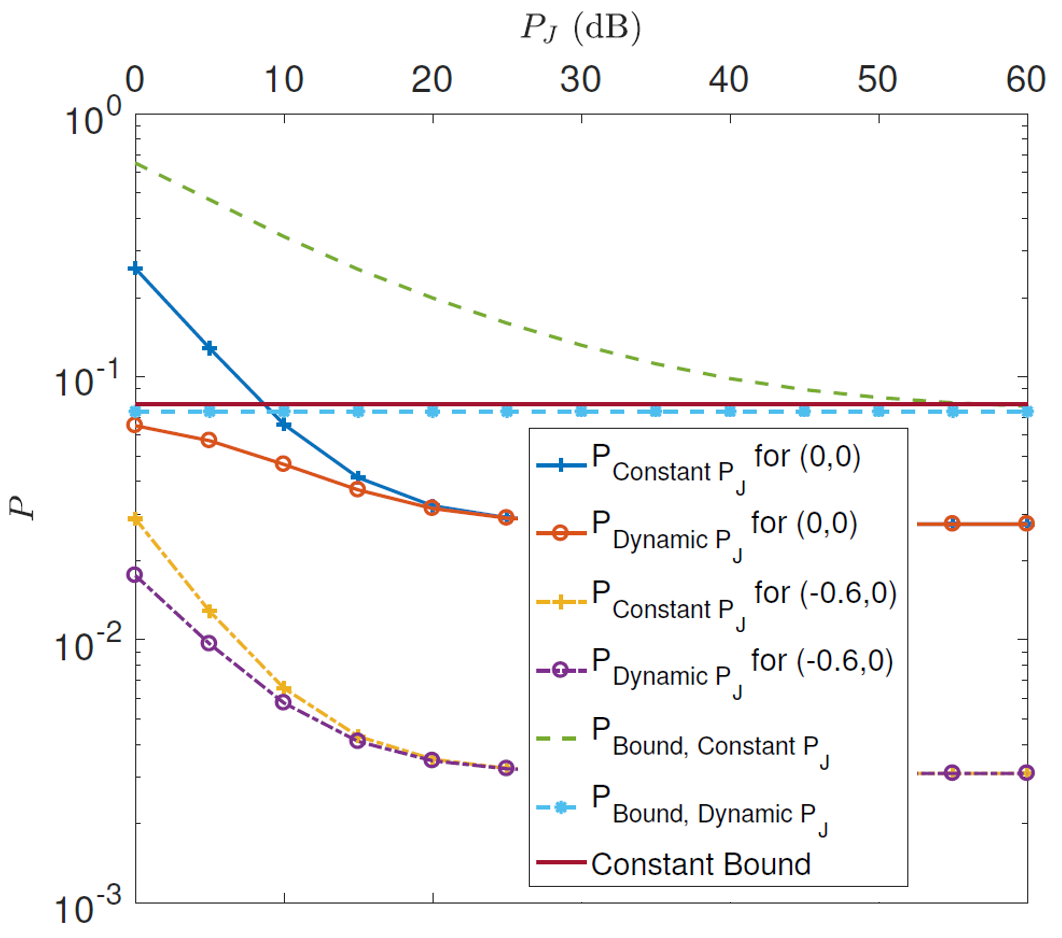}\\
  \caption{Probabilities shown in Properties  \ref{Prop_of_Prob_Sxy0_dynamic_PJ} and \ref{Prop_of_Prob_Sxy0_constant_PJ} where $\alpha=2$, $\rho=0.1$. The ``constant bound'' is $\frac{\pi \rho}{4}$. The meanings of other symbols should be self-evident. For location-dependent probabilities, we choose $(x,y)=(0,0)$ and $(x,y)=(-0.6,0)$.  }\label{prob_of_zero_secrecy_case_2_rho_01}
\end{figure}

Illustrated in Fig. \ref{prob_of_zero_secrecy_case_2_rho_01} are the probabilities shown in Properties  \ref{Prop_of_Prob_Sxy0_dynamic_PJ} and \ref{Prop_of_Prob_Sxy0_constant_PJ}. We see that the benefit from the semi-dynamic jamming power control
is significant when $P_J$ is small. But when $P_J$ is large, the benefit diminishes.

In scattering-rich environment, if Alice and Bob both move in random directions with a distance in the order of half-wavelength, all of $\tilde A$, $\tilde B_1$ and $\tilde B_2$ could change substantially and independently.
In MANET, Alice and Bob could move around and do not exchange any secret key till they find $\tilde A^2 > \rho^2 \tilde B_1 \tilde B_2$,  and only then they exchange secret keys with $P_J=P_J^*$. We call this a full-dynamic control of jamming power, for which the probability of zero secrecy is simply zero. Compared to the semi-dynamic control, the full-dynamic control has a larger latency and also potentially consumes more jamming power. Also note that the probability for the condition $\tilde A^2 > \rho^2 \tilde B_1 \tilde B_2$ to hold is $1-\mathcal{P}_1>1-\frac{\pi\rho}{4}$. So, if $\rho$ is small, Alice and Bob will find that condition  satisfied frequently.

\begin{figure}
  \centering
  \includegraphics[width=5cm]{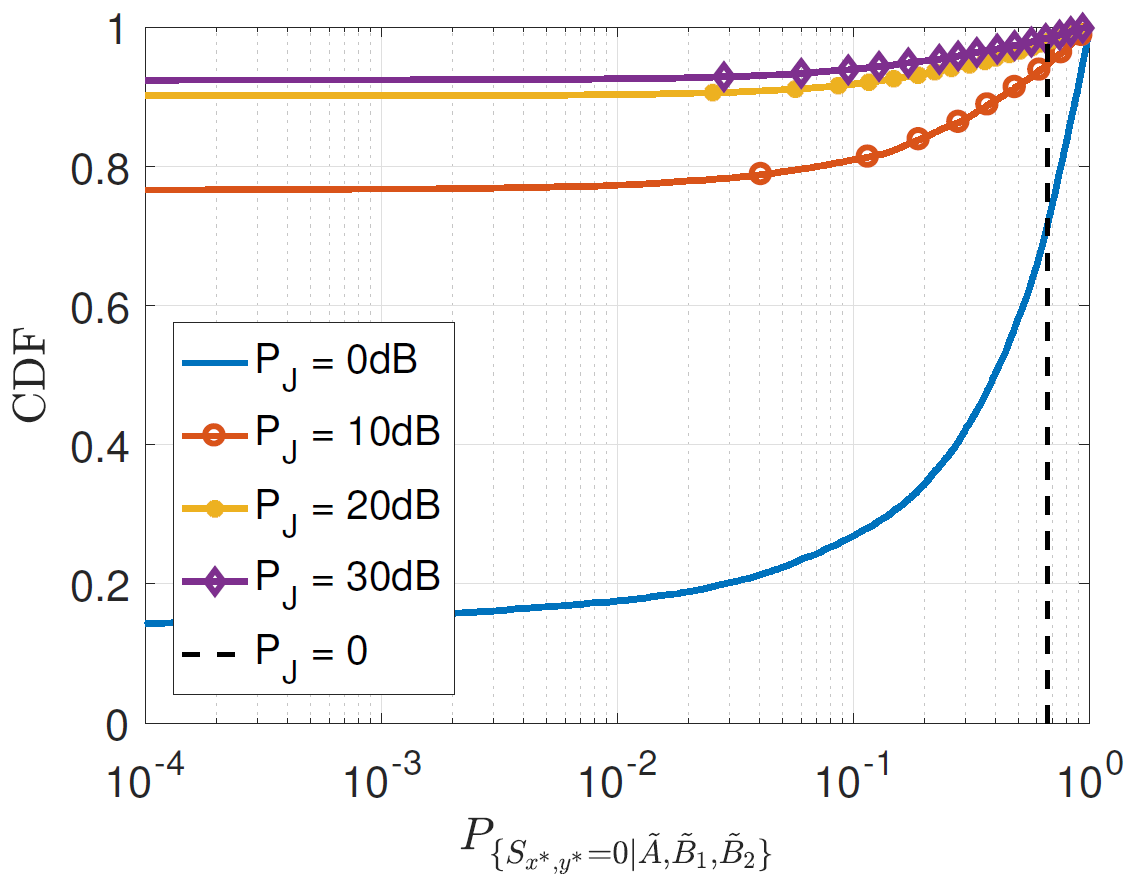}\\
  \caption{CDF of $\mathcal{P}_{\{S_{x,y}=0|\tilde A, \tilde B_1, \tilde B_2\}}$ with $\alpha=2$, $\rho=0.1$ and $(x,y)=(0,0)$. }\label{CDF_double_01_0_0}
\end{figure}

Alternatively, if $P_J$ is limited, Alice and Bob can wait till $\mathcal{P}_{\{S_{x,y}=0|\tilde A, \tilde B_1, \tilde B_2\}}$ shown in \eqref{eq:2D_integral} at a worst location of Eve is small enough (instead of zero). We call this a general-dynamic control.
Shown in Fig.  \ref{CDF_double_01_0_0}  are the CDFs of $\mathcal{P}_{\{S_{x,y}=0|\tilde A, \tilde B_1, \tilde B_2\}}$ at $x=y=0$, based on 10000 realizations of $\tilde A, \tilde B_1, \tilde B_2$. We see that even when $P_J$ is only zero dB (i.e., equal to the variance of the background noise), there is more than $10\%$ chance that $\mathcal{P}_{\{S_{x,y}=0|\tilde A, \tilde B_1, \tilde B_2\}}$ is less than $10^{-4}$. These CDFs should have a direct impact on the latency of the general-dynamic control.

\begin{Prop}\label{homogeneous_secrecy}
If $P_J=\sqrt{\frac{P_T}{\rho}}$ which is invariant to $(x,y)$, $\rho<\frac{\Delta^\alpha}{(1+\Delta)^\alpha}$, $1>d_A> \Delta$, $1>d_B> \Delta$, and $\Delta^\alpha \sqrt{\rho P_T}\gg 1$,  then with a high probability,
\begin{equation}\label{eq:Sxy2}
  S_{x,y}\approx \log_2\frac{\tilde A}{\rho\sqrt{\tilde B_1 \tilde B_2}}
\end{equation}
which is invariant to $(x,y)$. And under the above conditions,
 \begin{equation}\label{eq:Sxy2Bound}
   Prob\left \{S_{x,y}\leq s\right \}<\frac{2^s\rho\pi}{4}
 \end{equation}
\end{Prop}
  \begin{IEEEproof}
  Proof of \eqref{eq:Sxy2} is similar to that of Property \ref{constant_secrecy}. Proof of \eqref{eq:Sxy2Bound} is similar to that of \eqref{eq:P1}.
  \end{IEEEproof}

\begin{figure}
  \centering
  \includegraphics[width=5cm]{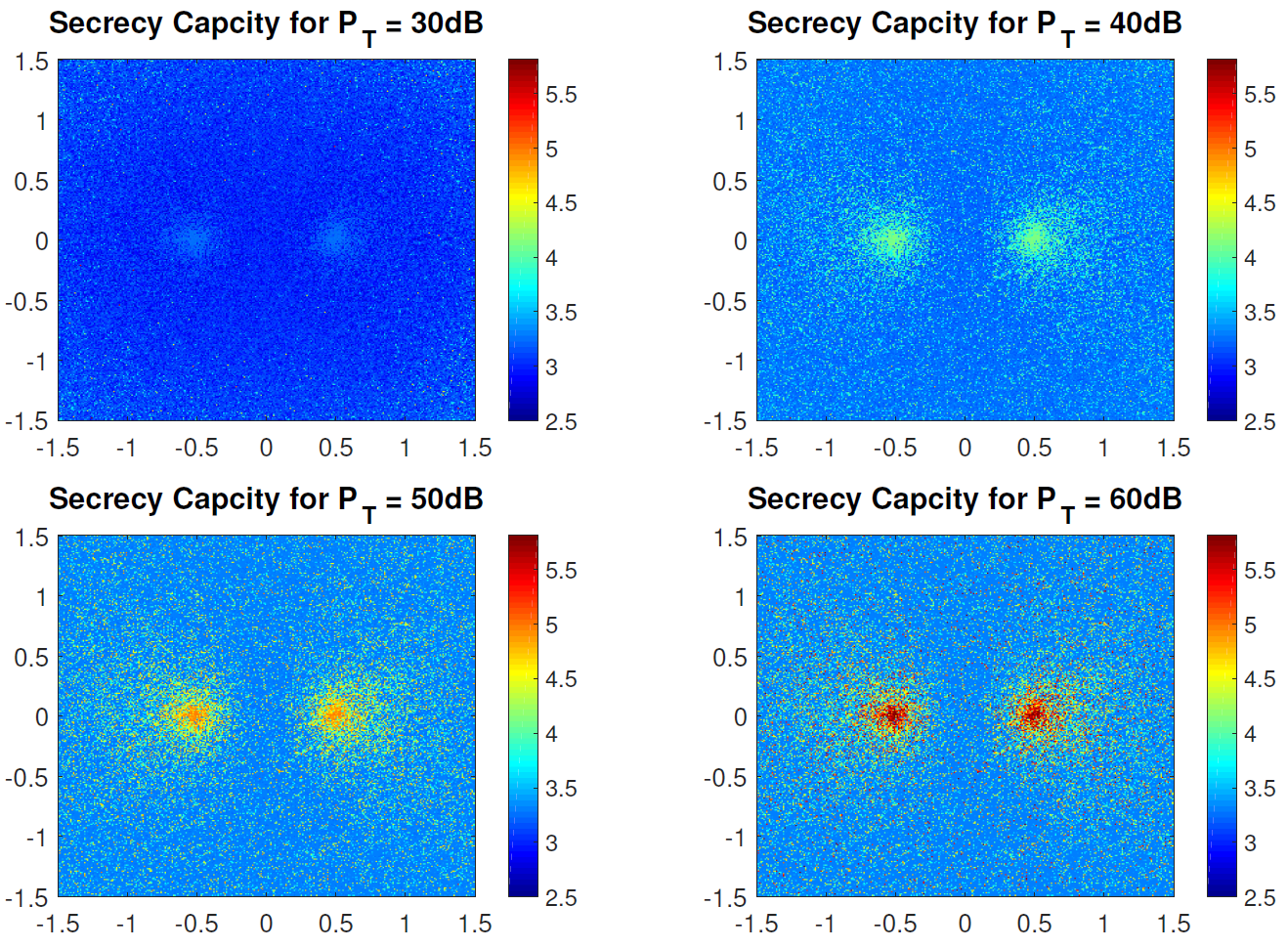}\\
  \caption{With small-scale fading and subject to $\tilde A=\tilde B_1 =\tilde B_2=1$, $S_{x,y}$ vs $(x,y)$. For each $(x,y)$, there is a new realization of $\tilde C$ and $\tilde D$. $\alpha=2$, $\rho=0.1$, $P_J=\sqrt{\frac{P_T}{\rho}}$. }\label{secrecy_capacity_dual_fading_01}
\end{figure}

\begin{figure}
  \centering
  \includegraphics[width=5cm]{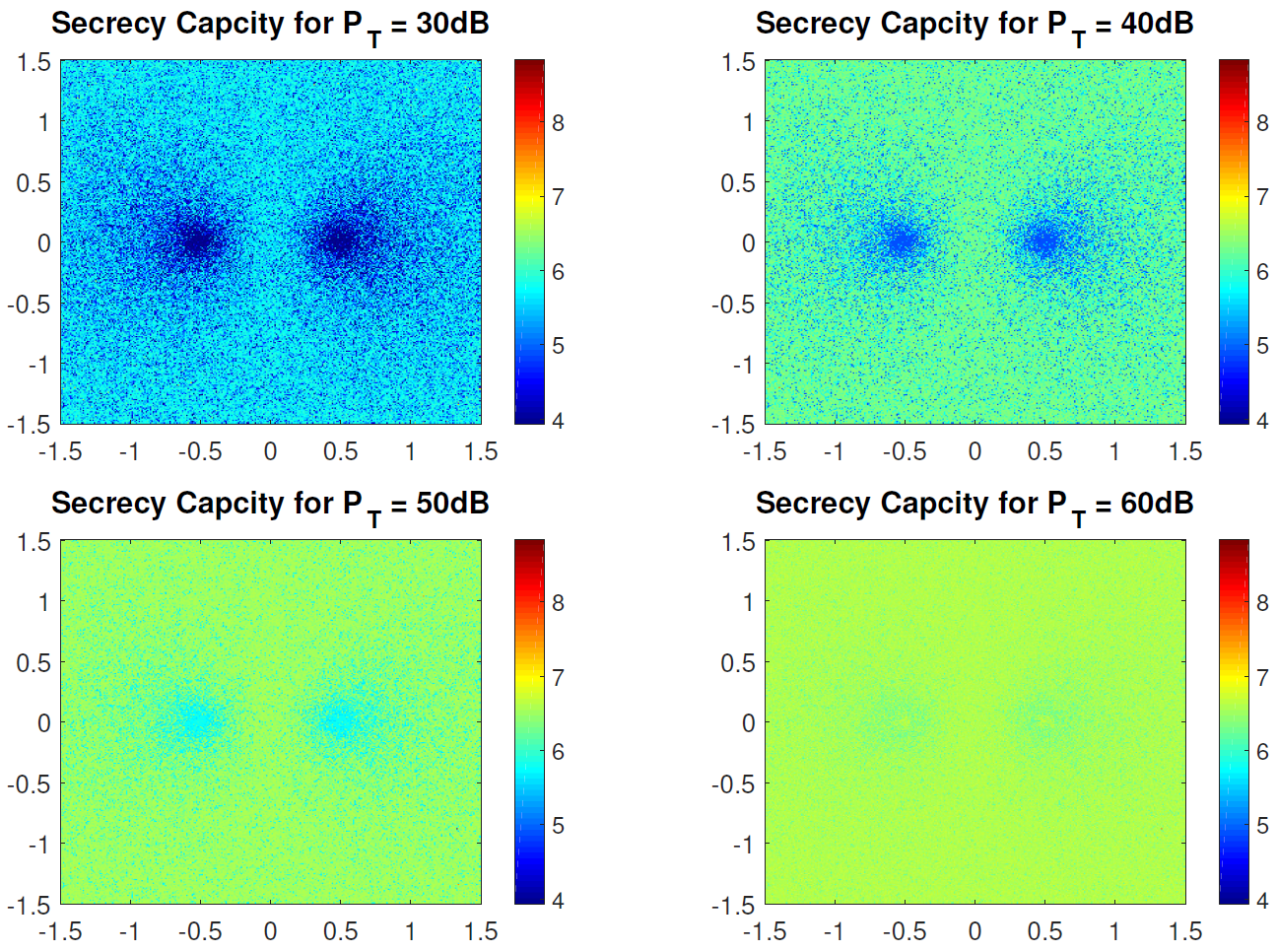}\\
  \caption{With small-scale fading and subject to $\tilde A=\tilde B_1 =\tilde B_2=1$, $S_{x,y}$ vs $(x,y)$. For each $(x,y)$, there is a new realization of $\tilde C$ and $\tilde D$. $\alpha=2$, $\rho=0.01$, $P_J=\sqrt{\frac{P_T}{\rho}}$. }\label{secrecy_capacity_dual_fading_001}
\end{figure}

The result \eqref{eq:Sxy2} says  that under a large $P_T$, a small $\rho$ and $P_J=\sqrt{\frac{P_T}{\rho}}$, there is a constant near-field secrecy capacity conditional upon $\tilde A$, $\tilde B_1$ and $\tilde B_2$. Illustrated in Figs. \ref{secrecy_capacity_dual_fading_01} and \ref{secrecy_capacity_dual_fading_001} is $S_{x,y}$ versus $(x,y)$ in the near-field subject to $\tilde A=\tilde B_1 =\tilde B_2=1$ where $\tilde C$ and $\tilde D$ change randomly and independently as $(x,y)$ changes (with step size equal to 0.01 in each direction). As predicted by Property \ref{homogeneous_secrecy}, we see that with $P_T=60dB$ and $\rho=0.01$, there is very little variation in the near-field distribution of $S_{x,y}$.


Since $S_{x,y}$ in \eqref{eq:Sxy2} still depends on $\tilde A$, $\tilde B_1$ and $\tilde B_2$,
 Alice and Bob could ideally wait till $\frac{\tilde A}{\sqrt{\tilde B_1 \tilde B_2}}$ is maximized before exchanging any keys. This is however not possible due to non-causality in knowing the future realizations of small-scale fading. In practice, Alice and Bob should only wait until $\log_2\frac{\tilde A}{\rho\sqrt{\tilde B_1 \tilde B_2}}$ is a large enough positive number. The result in \eqref{eq:Sxy2Bound} provides an upper bound on the probability that $\log_2\frac{\tilde A}{\rho\sqrt{\tilde B_1 \tilde B_2}}$ is no larger than a pre-specified value $s$ of secrecy.
 Obviously, this upper bound on the right side of \eqref{eq:Sxy2Bound} can be made small if $\rho$ is small enough. It is equivalent to write \eqref{eq:Sxy2Bound} as $Prob\left \{S_{x,y}> s\right \}>1-\frac{2^s\rho\pi}{4}$, which quantifies how to make the constant near-field secrecy capacity larger than a pre-specified value $s$ with high probability by choosing a small enough normalized self-interference channel gain $\rho$. Remember that under a fixed actual self-interference channel gain $\rho'$, the value of $\rho$ can be controlled by controlling the actual distance (or actual channel gain) between Alice and Bob. Such a control is feasible in mobile wireless networks.

\section{Final Remarks}\label{sec:final}
Full-duplex radio is an emerging wireless communication technology with many potential applications. This paper addresses its use for secure wireless communication. Unlike many previous works for this purpose, we have examined the fundamental properties of full-duplex radio for secure wireless communication where the legitimate users have virtually zero knowledge of the channel state information of eavesdroppers. Among the important findings are how the secrecy capacity of a link between two single-antenna radios is distributed in space with or without collusion among eavesdroppers, how the residual self-interference channel gain of full-duplex radio affects the distribution of the secrecy capacity, and how the small-scale fading affects the probabilities of zero (or positive) secrecy. All of the major findings have been quantified precisely and stated in a list of properties. Some of these properties might not be surprising to some experts in this area. But to our knowledge, none of these properties is  readily available elsewhere (from other authors). Indeed we hope that all of them are original and insightful. We also hope that this work will be useful for helping real-world applications and more importantly for inspiring further research in understanding the limits and potentials of full-duplex radio for secure wireless communication in more advanced network settings.

\end{document}